\documentclass[aps,prb,groupedaddress,reprint,notitlepage]{revtex4-1}
\usepackage{amsmath}
\usepackage{graphicx}
\usepackage{bm}
\usepackage{float}
\usepackage{appendix}
\newcommand{\ve}{\varepsilon}
\renewcommand{\vec}[1]{\mathbf{#1}}

\newcommand{\mctwo}{Department of Microtechnology and Nanoscience -- MC2,
Chalmers University of Technology, SE-41296 Gothenburg, Sweden}

\begin{document}

\title{Signatures of van der Waals binding: A coupling-constant scaling analysis}

\author{Yang Jiao}
\email[]{yang.jiao@chalmers.se}
\affiliation{\mctwo}
\author{Elsebeth Schr{\"o}der}
\affiliation{\mctwo}
\author{Per Hyldgaard}
\affiliation{\mctwo}

\date{\today}

\begin{abstract}
The van der Waals (vdW)  density functional (vdW-DF) method [ROPP 78, 066501 (2015)] 
describes dispersion or vdW binding by tracking the effects of an electrodynamic 
coupling  among pairs of electrons and their associated exchange-correlation holes. 
This is done in a
nonlocal-correlation energy term $E_{\rm c}^{\rm nl}$, which permits density functional 
theory calculation in the Kohn-Sham scheme. However, to map the nature of 
vdW forces in a fully interacting materials system, it is necessary to also account for 
associated kinetic-correlation energy effects. Here, we present a 
coupling-constant scaling analysis which permits us to compute the 
kinetic-correlation energy $T_{\rm c}^{\rm nl}$ that 
is specific to the vdW-DF account of nonlocal correlations.
We thus provide a more complete spatially-resolved analysis of the 
electrodynamical-coupling nature of nonlocal-correlation binding, including 
vdW attraction, in both covalently and non-covalently bonded systems. We find 
that kinetic-correlation energy effects play a significant role in the account 
of vdW or dispersion interactions among molecules. Furthermore, our 
mapping shows that the total nonlocal-correlation binding is concentrated to pockets 
in the sparse electron distribution located between the material fragments. 

\end{abstract}

\maketitle

\section{Introduction}

Many-body effects are essential for an accurate description of materials bonds.
Such correlation effects must be accurately reflected in density functional 
theory (DFT) as we seek approximate evaluations of ground-state expectation values, 
$\langle \hat{T} \rangle$ and $\langle \hat{V} \rangle$, of operators for the 
kinetic energy $\hat{T}$ and for the electron-electron interaction $\hat{V}$. This is
clear, for example, because dispersion or van der Waals (vdW) interactions arise from an 
electrodynamical coupling among collective excitations.\cite{jerry65,lape77,ra,ma,lavo87,hybesc14}  
In the Kohn-Sham (KS) scheme\cite{kosh65} for efficient, in principle, exact DFT calculations, 
we handle all many-body effects by a trick. We focus on an independent-particle approximation, 
the KS kinetic-energy term $T_{\rm KS}$, while embedding the difference, termed the
kinetic-correlation energy,
\begin{equation} 
    T_{\rm c} = \langle \hat{T} \rangle - T_{\rm KS} \, ,
    \label{eq:kincordef}
\end{equation}
in the exchange-correlation (XC) energy, $E_{\rm xc}$. However, the
kinetic-correlation energy, Eq.\ (\ref{eq:kincordef}), can  still
be unmasked as a functional $T_c[n]$ of the electron density $n(\vec{r})$
using a formally exact scaling analysis.\cite{Levy85,Levy91,Gorling93,Burke97}
An evaluation of $E_{\rm xc}[n]-T_{\rm c}[n]$ is equivalent to correcting $\langle \hat{V} \rangle$ 
beyond the Hartree approximation\cite{Levy85} and thus allows an exploration
of many-electron interaction effects.

The van der Waals (vdW) density functional (vdW-DF) method for general-purpose
DFT calculations relies on truly nonlocal formulations $E_{\rm xc}[n]$, Refs.\ \onlinecite{anlalu96,ryluladi00,rydberg03p126402,Dion,thonhauser,lee10p081101,behy14,Thonhauser_2015:spin_signature,Berland_2015:van_waals}. The vdW-DF functional design can be seen as a systematic 
extension of the local density approximations (LDA) and of the generalized gradient 
approximation (GGA). In its original and most commonly used form,\cite{Dion,behy14} 
it relies on the same many-body perturbation theory analysis\cite{lavo87,thonhauser} 
that underpins the formulations of PBE\cite{pebuer96} and PBEsol,\cite{PBEsol} and it 
adheres to the same fundamental principle, that physics constraints, including 
charge\cite{Dion} and current\cite{Dion,hybesc14} conservation, should guide the XC 
functional design.\cite{rydberg03p126402,rydbergthesis,bearcoleluscthhy14} 
However, as part of a rationale for constraint-based GGA, Langreth 
and Vosko showed that a gradient-corrected formulation of correlation cannot 
naturally account for vdW interactions.\cite{lavo87} The vdW-DF method overcomes
that limitation, noting that electrons and their associated GGA-type XC holes 
themselves form dipole systems with internal dynamics.\cite{ma,ra,anlalu96,becke05p154101,becke07p154108,hybesc14}
The vdW-DF method tracks screening effects produced by the mutual electrodynamical
coupling of such virtual dipoles. It thus extends GGA within the vdW-DF framework, 
capturing screened dispersion binding\cite{ma,ra} by summing coupling-induced shifts in the 
collective plasmon excitations.\cite{jerry65,hybesc14}

The vdW-DF method is computationally efficient since the dispersion-energy 
gains\cite{hybesc14} are evaluated in a truly nonlocal-correlation energy 
term $E_{\rm c}^{\rm nl}$ that is an explicit functional of the density.\cite{Dion,roso09} 
This is done by using the adiabatic-connection formula\cite{lape75,gulu76,lape77} 
(ACF) for the exact XC functional to define an effective dielectric 
function $\kappa$,\cite{Dion,hybesc14,Berland_2015:van_waals}  and by 
expanding $\kappa$ in terms of a plasmon-pole approximation that reflects 
the response corresponding to an internal semilocal functional\cite{rydbergthesis,Dion,lee10p081101,hybesc14} 
$E_{\rm xc}^{\rm in}$. In the original general-geometry vdW-DF\cite{Dion}  
and in the recent consistent-exchange vdW-DF-cx\cite{behy14} 
formulations,\footnote{In the case of 
vdW-DF2\cite{lee10p081101} by a formulation that reflects an exchange-scaling to the 
high-density limit.} this internal function comprises LDA with gradient corrections defined by analysis of screened exchange.\cite{lavo87,Dion,thonhauser,Thonhauser_2015:spin_signature}
The total functional specification\cite{Dion,behy14}
\begin{equation}
E_{\rm xc}^{\rm vdW-DF}[n] = E_{\rm xc}^{\rm in} + E_{\rm c}^{\rm nl} + \delta E_{\rm x}^0 \, ,
\label{eq:fullvdWDF}
\end{equation}
generally also contains a cross-over term $\delta E_{\rm x}^0$ that contains nothing but 
gradient-corrected exchange.\cite{bearcoleluscthhy14,Berland_2015:van_waals} The
total exchange functional $E_{\rm x}$ is semilocal; the correlation part of the functional,
$E_{\rm c}$, comprises LDA correlation $E_{\rm c}^{\rm LDA}$ (from $E_{\rm xc}^{\rm in}$) and
$E_{\rm c}^{\rm nl}$.

An elegant illustration of the many-body physics nature of vdW binding
can be obtained by computing the spatially resolved component\cite{callsen12p085439,rationalevdwdfBlugel12}
\begin{equation}
e_{\rm c}^{\rm nl}[n](\vec{r}) =  \frac{n(\vec{r})}{2} \int_{\vec{r'}} 
\Phi(n(\vec{r}),\nabla n(\vec{r}),n(\vec{r'}),\nabla n(\vec{r'})) n(\vec{r'}) \, ,
\label{eq:locecnl}
\end{equation}
of the total vdW-DF nonlocal-correlation energy
\begin{equation}
E_{\rm c}^{\rm nl}[n] =  \int_{\vec{r}} \, e_{\rm c}^{\rm nl}[n](\vec{r})\, .
\label{eq:ecnlresolve}
\end{equation} 
The spatial resolution Eq.\ (\ref{eq:ecnlresolve}) is a natural extension
of how we normally analyze total-energy contributions arising from the 
semilocal components of the XC energy.\cite{lape77,pebuer96,Burke97,Dion}
The spatially resolved energy Eq.\ (\ref{eq:locecnl}) is 
given by the vdW-DF 
kernel\cite{Dion,dionerratum} $\Phi$ for which there exist both formal 
analysis\cite{thonhauser} and an efficient evaluation scheme.\cite{roso09}
With Eq.\ (\ref{eq:locecnl}) one can track and understand 
binding-induced changes $\Delta e_{\rm c}^{\rm nl}(\vec{r})$, 
for example, for benzene adsorption on graphene.\cite{rationalevdwdfBlugel12} 
The mapping confirms that the dominant contributions to the vdW binding arise
in the regions of sparse\cite{langrethjpcm2009} (but not vanishingly low) electron 
density between molecules and surfaces.\cite{rydberg03p126402,kleis08p205422,berland10p134705,behy13,hybesc14}

In this paper, we seek a characterization of electron-electron interaction effects 
that underpin vdW attraction between molecules. Many computational descriptions of 
the vdW attraction build on a discussion of the electron response and dielectric 
function in the physical, fully interacting system,\cite{li56,jerry65,zarembakohn1976,zarembakohn1977,harrisnordlander1984,huhyrolu01,ra,ma,anlalu96,ryluladi00,rydberg03p126402,kleis07p100201,grimme1,grimme2,grimme3,becke05p154101,becke07p154108,silvestrelli08p53002,ts09,vv10,ts2,ts-mbd} although the actual 
response behavior is sometimes approximated by an independent-particle description. The ACF specifies the 
exact XC functional as an average over the electron response, denoted $\chi_{\lambda}$, that reflects a 
ramping ($0 <\lambda < 1$) of assumed electron-electron interaction strengths, $\hat{V}_{\lambda}=\lambda 
\hat{V}$.\cite{lape75,gulu76,lape77,pebuwa96,Burke97} This ACF view is explicitly maintained in vdW 
density functionals\cite{luanetal95,dobdint96,Dion,bearcoleluscthhy14,Berland_2015:van_waals,Thonhauser_2015:spin_signature} 
which track the vdW binding produced by plasmon-energy shifts\cite{jerry65,ra} in a KS framework.\cite{lape77,hybesc14} 
However, the electrodynamical-coupling mechanism for vdW attraction\cite{ma,ra} is at work in the physical 
system, i.e., at full coupling-constant strength $\lambda=1$. For a more complete mapping of the
nature of vdW attraction,\cite{ra} we therefore seek to (a) compute an XC energy, denoted 
$E_{{\rm xc},\lambda=1}[n]$, that instead reflects the physical response $\chi_{\lambda=1}$, and 
(b) extract and study the component, denoted $E_{{\rm c},\lambda=1}^{\rm nl}[n]$, 
that corresponds to nonlocal-correlation effects in $\chi_{\lambda=1}$. 

Our central observation is that such information is directly available from the vdW-DF 
functional form, Eq.\ (\ref{eq:fullvdWDF}), by applying the formally exact coupling-constant scaling 
analysis\cite{Levy85,Levy91,Levy95chapter} on the vdW-DF method.  The formal analysis rests on density scaling, which provides a complete specification of the would-be XC energy $E_{{\rm xc},\lambda}[n]$ that reflects the response function $\chi_{\lambda}$ assuming only that the $\lambda$-averaged response defines the specific
$E_{\rm xc}[n]$ form; the analysis can be made for a given problem once we know the self-consistent solution density $n(\vec{r})$. 
We present details of how to extend the scaling analysis from semilocal functionals\cite{Gorling93,Perdew96,Burke97,Ernzerhof97} to the truly nonlocal-correlation 
term $E_{\rm c}^{\rm nl}[n]$ of the vdW-DF method. 

We note that the formal scaling analysis permits calculations of the kinetic-correlation 
energy,\cite{Levy85} Eq.\ (\ref{eq:kincordef}). For practical calculations, we present a code, 
termed \textsc{ppACF}, that computes the component
\begin{equation}
T_{\rm c}^{\rm nl}[n] = \int_{\vec{r}}\, t_{\rm c}^{\rm nl}[n](\vec{r}) \, ,
\label{eq:tcnlresolve}
\end{equation}
which is specific to $E_{\rm c}^{\rm nl}[n]$. Eq.\ (\ref{eq:tcnlresolve})
is also combined with the known coupling-constant scaling analysis for LDA
correlation,\cite{Levy85,Levy91,Gorling93,Levy95chapter} for a full specification of the kinetic-correlation energy
\begin{equation}
T_{\rm c}[n] = \int_{\vec{r}} \, t_{\rm c}[n](\vec{r}) \, .
\label{eq:tcresolve}
\end{equation}
Finally we rely on the formal equivalence\cite{Levy85}
\begin{equation}
    E_{{\rm xc},\lambda=1}[n] \equiv E_{\rm xc}[n]-T_{\rm c}[n] \, ,
    \label{eq:fullExcdef}
\end{equation}
to extract a representation
\begin{equation}
    E_{{\rm c},\lambda=1}^{\rm nl}[n] \equiv E_{\rm c}^{\rm nl}[n]-T_{\rm c}^{\rm nl}[n] \, ,
    \label{eq:fullEcnldef}
\end{equation}
of the mutual plasmon electrodynamical coupling in the physical 
system.\cite{jerry65,ra,ma,hybesc14} 

As implied in Eqs. (\ref{eq:tcnlresolve}) and (\ref{eq:tcresolve}), the code also gives us
access to spatially resolved kinetic-correlation energies, $t_{\rm c}^{\rm nl}[n](\vec{r})$
and $t_{\rm c}[n](\vec{r})$, that are consistent with Eqs.\ (\ref{eq:locecnl}) and (\ref{eq:ecnlresolve})
and with the standard resolution of XC energy contributions. Using \textsc{ppACF}, we can thus
compute and discuss the nature of binding-induced changes 
$\Delta t_{\rm c}(\vec{r})$, $\Delta t_{\rm c}^{\rm nl}(\vec{r})$, 
and $\Delta e_{\rm c}^{\rm nl}(\vec{r})-\Delta t_{\rm c}^{\rm nl}(\vec{r})$,
in the spatially resolved descriptions. Our \textsc{ppACF} code can provide 
this analysis for most versions or variants of the vdW-DF method.\cite{Dion,lee10p081101,cooper10p161104,optx,vdwsolids,behy14,hamada14,Berland_2015:van_waals} 
Here we work with the  consistent-exchange vdW-DF-cx formulation,\cite{behy14,bearcoleluscthhy14} 
which can effectively be seen as a mean-value  evaluation of the ACF.\cite{hybesc14} 

We find that $\Delta t_{\rm c}(\vec{r})$ and $\Delta e_{\rm c}^{\rm nl}(\vec{r})$ both 
contain signatures of directed binding: the dominant binding contributions are
channeled into pockets. Also, the signatures in $\Delta t_{\rm c}^{\rm nl}(\vec{r})$ and
in $\Delta e_{\rm c}^{\rm nl}(\vec{r})$ typically mirror each other, 
up to a sign. This means that the concentration of vdW bonding is 
further enhanced in the contribution $\Delta e_{\rm c}^{\rm nl}(\vec{r}) - 
\Delta t_{\rm c}^{\rm nl}(\vec{r})$ that characterizes the electrodynamical
coupling mechanism behind the vdW attraction.\cite{jerry65,ma,ra,hybesc14}

Overall, our results show that there is an important kinetic-energy nature of 
vdW binding and confirm that the density tails, rather than the atomic centers, 
play the decisive  role in setting dispersion forces at binding separations.\cite{rydberg03p126402,kleis08p205422,berland10p134705,berland11p1800,lee11p193408,lee12p104102,rationalevdwdfBlugel12,behy13,behy14,hybesc14} Our results also 
suggest that there exists an orbital-like structure of dispersion binding, although
much weaker than in chemical bonds and originating in different mechanisms.\cite{ma,thonhauser,hybesc14} 
This observation could be useful for qualitative discussions of the nature and
variation in vdW forces in materials.

The rest of this paper is organized as follows.
Section II details the coupling constant analysis of the vdW-DF method. Section
III provides computational details. In Sec.\ IV we document
signatures of the vdW attraction in both noncovalent and covalent 
molecular binding. Section V contains a summary and discussion.
The paper has one appendix.

\section{Theory}

A systematic theory characterization of the screened response in a 
homogeneous and weakly perturbed electron 
gas\cite{lape75,gulu76,lape77,lape80,lameprl1981,lavo87,lavo90}
has led to the definition of a range of successful constraint-based 
functionals for the XC energy $E_{\rm xc}$ and broad use of DFT.
We use $\hat{V}$ to denote the full electron-electron interaction.
We consider the density changes $\delta n$ produced by an external 
field $\delta \Phi_{\rm ext}$, and compute the electron-gas density 
response $\chi_\lambda=\delta n/\delta \Phi_{\rm ext}$ 
as a function of the assumed coupling constant $\lambda$ for 
an adiabatic turn on of the many particle interaction, $V_\lambda = 
\lambda V$. The exact XC energy is given by the ACF,
\begin{equation}
E_{\rm xc} = - \int_0^\infty \, \frac{du}{2\pi} \, \hbox{Tr} \{ \chi_\lambda(iu) V \} 
-E_{\rm self}\, , 
\label{eq:ACF}
\end{equation}
which links $\lambda$, the (complex) frequency $iu$, and spatial variations in
the response function $\chi_\lambda$ to the XC energy. 
We use $\hat{n}(\vec{r})$ to denote the density operator, and the last term
of Eq.\ (\ref{eq:ACF}) is the electron self energy 
$E_{\rm self}  =  \hbox{Tr} \{ \hat{n} V\}/2.$

The exact XC energy can be recast as an electrostatic interaction\cite{lape75,gulu76,lape77}
\begin{equation}
E_{\rm xc} = \frac{1}{2} \int_\vec{r}\int_{\vec{r'}} \frac{n(\vec{r})
n_{\rm xc}(\vec{r}; \vec{r'})}{|\vec{r}-\vec{r'}|}    
\label{eq:XCrecast}
\end{equation}
between the electrons and associated, so-called, XC holes $n_{\rm xc}(\vec{r};\vec{r'})$.
The XC hole  reflects a $\lambda$ average of the response $\chi_\lambda$. An emphasis 
on the assumed plasmon-nature of the electron-response, a reliance on formal many-body perturbation 
theory, and the imposing of additional physics constraints, such as charge conservation 
of the XC hole, has led to formulations of LDA,\cite{helujpc1971,pewa92} 
of the PBE  and PBEsol versions of GGAs,\cite{pebuer96,PBEsol} and of
the vdW-DF method.\cite{Dion,thonhauser,lee10p081101,bearcoleluscthhy14,hybesc14,Berland_2015:van_waals}

At any given coupling constant $\lambda$, the response function
defines an approximation for the exchange-correlation hole
\begin{equation}
n_{{\rm xc},\lambda}(\mathbf{r},\mathbf{r'}=\mathbf{r}+\mathbf{w}) = 
-\frac{2}{n(\vec{r})}\, \int_0^\infty \, \frac{du}{2\pi} \,
\chi_\lambda(\mathbf{r},\mathbf{r'}; iu) -\delta(\vec{w})\, .
\label{eq:lambdahole}
\end{equation}
The actual XC hole then emerges simply as an average,
\begin{equation}
n_{\rm xc}=\int_0^1 n_{{\rm xc},\lambda} \, d\lambda \, .
\end{equation}
Using Eq.\ (\ref{eq:lambdahole}) it is meaningful to define and discuss also
the coupling-constant dependence of the XC functional:
\begin{equation}
E_{{\rm xc},\lambda} \equiv \frac{1}{2} \int_{\mathbf{r}}\int_{\mathbf{r'}}
\frac{n(\mathbf{r}) \, n_{{\rm xc},\lambda}(\mathbf{r},\mathbf{r'}) }{|\mathbf{r}-\mathbf{r}'|} \, .
\label{eq:ExcVar}
\end{equation}
Same as for the holes, the actual functional, Eq.\ (\ref{eq:XCrecast}), is given by an average 
over $0<\lambda < 1$, 
\begin{equation}
E_{\rm xc}[n] = \int_0^1 \, d\lambda \, E_{{\rm xc},\lambda}[n]\, .
\label{eq:lambdaresolve}
\end{equation}

The behavior of $E_{{\rm xc},\lambda}$ is exclusively set by exchange effects at $\lambda=0$. This follows because exchange reflects
an independent-particle behavior and, unlike correlation, it is
independent of $\lambda$. At the other physical limit, the plasmon character can be expected to dominate in the response. One therefore also expects that $E_{{\rm xc},\lambda}$
becomes accurate at $\lambda \to 1 $ if Eq.\ (\ref{eq:ExcVar}) reflects a plasmon-based analysis of electron response, for example, as used the early LDA formulations \cite{helujpc1971,gulu76}, in the constraint-based GGAs \cite{pebuer96,PBEsol}, and in vdW-DF-cx \cite{behy14}.

It is instructive to  split the XC hole into exchange and correlation components
\begin{equation}
    n_{{\rm xc},\lambda}(\vec{r},\vec{r}')
    = n_{\rm x}(\vec{r},\vec{r}') 
      + n_{{\rm c},\lambda}(\vec{r},\vec{r}') \, ,
      \label{eq:holesplit}
\end{equation}
and to define (at every $\lambda$) a spatially resolved
correlation term
\begin{equation}
    e_{{\rm c},\lambda}[n](\vec{r})= \frac{n(\vec{r})}{2} 
    \int_{\vec{r'}}
    \frac {n_{{\rm c},\lambda}(\vec{r},\vec{r}')}
    {|\vec{r}-\vec{r'}|} \, .
\end{equation}
This term provides a mapping of the total  
correlation effects at $\lambda$:  
\begin{equation}
    E_{{\rm c},\lambda}[n] = 
    \int_{\vec{r}} e_{{\rm c},\lambda}(\vec{r}) \, .
\end{equation}
Also, there exists a coupling-constant scaling
analysis for LDA correlation\cite{Levy95chapter,Gorling93,Burke97}
$E_{{\rm c},\lambda}^{\rm LDA}$ with spatial resolution
\begin{equation}
E_{{\rm c},\lambda}^{\rm LDA}[n] = \int_{\vec{r}} 
    e_{{\rm c},\lambda}^{\rm LDA}[n](\vec{r}) \, .
    \label{eq:LDAcorscale}
\end{equation}
Accordingly we isolate a spatially resolved 
nonlocal-correlation energy
\begin{equation}
    e_{{\rm c},\lambda}^{\rm nl}[n](\vec{r})
    = e_{{\rm c},\lambda}[n](\vec{r}) - 
    e_{{\rm c},\lambda}^{\rm LDA}[n](\vec{r}) \, ,
    \label{eq:ecnl-lambdadf}
\end{equation}
corresponding the coupling-constant scaling of the
total nonlocal-correlation energy
\begin{equation}
E_{{\rm c},\lambda}^{\rm nl}[n] = \int_{\vec{r}} e_{{\rm c},\lambda}^{\rm nl}[n](\vec{r}) \, .
\label{eq:energydensity}
\end{equation}
Equation\ (\ref{eq:locecnl}) is the coupling-constant integral of $e_{{\rm c},\lambda}^{\rm nl}[n](\vec{r})$.

To map the electrodynamical-coupling nature of vdW attraction, we seek to compute binding-induced changes $\Delta e_{{\rm c},\lambda=1}^{\rm nl}[n](\vec{r})$.

\subsection{Density scaling in the exact XC energy}

Coupling-constant scaling analysis\cite{Levy85} is a natural
tool for exploring the nature of both exchange-based GGAs\cite{Burke97,Ernzerhof97}
and of vdW-DF-cx.  For any given solution density $n(\vec{r})$, we define
a rescaled density
\begin{equation}
n(\mathbf{r}) \to n_{1/\lambda}(\mathbf{r}) \equiv n(\mathbf{r}/\lambda)/\lambda^3 \, ,
\label{eq:densscale}
\end{equation}
and resolve Eq.\ (\ref{eq:lambdaresolve}) into $\lambda$-specific contributions using
the exact result\cite{Levy85,Burke97,Ernzerhof97}
\begin{equation}
 E_{{\rm xc},\lambda}[n] = \frac{d}{d\lambda} \left\{ \lambda^2 E_{{\rm xc}}[n_{1/\lambda}] \right\} \, .
\label{eq:scaleExc}
\end{equation}
Since there is no $\lambda$ dependence for exchange, we can recast 
Eq.\ (\ref{eq:scaleExc}) using the correlation-energy density:
\begin{equation}
 e_{{\rm c},\lambda}[n](\vec{r}) = 
 \frac{d}{d\lambda} \left\{ \lambda^2 e_{{\rm c}}[n_{1/\lambda}](\vec{r}) \right\} \, .
\label{eq:scaleEcdensity}
\end{equation}
The scaling results for $E_{{\rm xc},\lambda}[n]$ and $e_{{\rm c},\lambda}[n]$ can be directly applied 
to individual components of the XC functional (as they are linear in the functional expression).

The scaling results, Eqs.\ (\ref{eq:scaleExc}) and (\ref{eq:scaleEcdensity}), reflect properties 
of the $\chi_{\lambda}$ approximations that are  implicitly made in crafting the  PBE and vdW-DF-cx functionals. The existence of a well-understood coupling-constant scaling has been used to rationalize the formulation of the PBE0 hybrid\cite{PBE0} based on PBE.\cite{Burke97,Ernzerhof97} 
Noting that a similar rationale exists for the coupling constant scaling of vdW-DF-cx, some of us have recently motivated the introduction of correspondingly defined vdW-DF hybrids, 
including vdW-DF-cx0, which replace the 
vdW-DF-cx exchange component with a fraction 
of Fock exchange.\cite{DFcx02017}

The scaling results, Eqs.\ (\ref{eq:scaleExc}) and (\ref{eq:scaleEcdensity}),
follow from an
analysis of the many-particle wavefunction ground-state solution $\Psi_n^{{\rm min},\lambda}$ corresponding to a specific density $n$ 
and a specific strength $\lambda V$ of the electron-electron interaction. 
The detailed arguments are given elsewhere; For completeness, we include
a renormalization-type argument for this observation in the appendix. 
Here we simply note that the wavefunctions solving the  
Hamiltonian $\Hat{H}=\hat{T} + \lambda V+ V_{\rm ext}$ themselves 
scale according to 
\begin{equation}
\Psi_n^{{\rm min},\lambda}(\mathbf{r}_1,\ldots,\mathbf{r}_N) = \lambda^{3N/2}
\Psi_{n_{1/\lambda}}(\lambda\mathbf{r}_1,\dots,\lambda\mathbf{r}_N) \, ,
\label{eq:wavescale}
\end{equation}
and that this formal equivalence is sufficient to establish the $\lambda$
scaling.\cite{Levy85,Burke97}

\subsection{Access to the kinetic-correlation energy}

Below we drop explicit references to the density functional
nature when working with spatially-resolved energy contributions
such as $e_{{\rm c},\lambda}^{\rm nl}(\vec{r})$  and
$e_{{\rm c}}^{\rm nl}(\vec{r})$, except when specifically needed
for the discussion.

In the KS scheme \cite{kosh65}, we nominally focus on computing the so-called KS kinetic energy\footnote{In discussions of DFT, $T_{\rm KS}$ is sometimes called the single-particle kinetic energy. We prefer the Kohn-Sham label as $\hat{T}$ is always a
single-particle operator.} from single-particle expectation values 
\begin{eqnarray}
\langle \phi_i | \hat{T} | \phi_i \rangle &  = &  \int_{\vec{r}} \tilde{t}_{i}(\vec{r}) \,, 
\label{eq:TiDEF}\\
\tilde{t}_{i} (\vec{r}) & = & - \frac{1}{2}  \phi_i^{*}(\vec{r}) 
\nabla^2 \phi_i(\vec{r}) \, ,
\label{eq:TksdefIntg2}
\end{eqnarray}
for occupied orbitals $\phi_i(\vec{r})$. As in the Quantum-Espresso 
package,\cite{QE} we compute the KS kinetic energy as a spatial integration
\begin{equation}
  T_{\rm KS}[n] = \int_{\vec{r}} t_{\rm KS}(\vec{r}) \,,
  \label{eq:TKSdef}
\end{equation}
over positive definite contributions 
\begin{equation}
t_{\rm KS}(\vec{r}) =  \frac{1}{2} \sum_i^{\rm occ} |\nabla \phi_i(\vec{r})|^2 \, ,
\label{eq:TksdefIntg}
\end{equation}
defined by the set of occupied orbitals. This representation of the KS kinetic energy 
is simply related to the summation $\tilde{t}_{\rm occ}(\vec{r}) \equiv \Sigma_i^{\rm occ} \tilde{t}_i(\vec{r})$ over single-particle contributions, Eq.\ (\ref{eq:TksdefIntg2}). 
The descriptions differ only in the inclusion of an Poisson-type term
\begin{equation}
    t_{\rm KS}(\vec{r}) = \frac{1}{4} \nabla^2 n(\vec{r}) + 
    \hbox{Re} \{\tilde{t}_{\rm occ}(\vec{r}) \} \, ,
    \label{eq:tkschanges}
\end{equation}
and give the same total KS kinetic energy, Eq.\ (\ref{eq:TKSdef}), upon spatial 
integration.

We typically compute DFT energies $E^{\rm A/B/AB}_{\rm DFT}$ of combined
systems `AB' and of the relevant fragments, `A' or `B', to understand 
binding $\Delta E_{\rm DFT} = E^{\rm A}_{\rm DFT} +  
E^{\rm B}_{\rm DFT} - E^{\rm AB}_{\rm DFT}$ (with suitable adjustments
in the case of related problems such as material cohesion). The mean-field
electrostatic energy among electrons, that is, the Hartree term
\begin{equation}
 U[n] = \frac{1}{2} \int_\mathbf{r}\int_\mathbf{r'} 
 \frac{n(\mathbf{r})\, n(\mathbf{r'})}{|\mathbf{r}-\mathbf{r'}|} \, ,
 \label{eq:Udef}
\end{equation}
is one important contribution as it approximates $\langle \hat{V} \rangle$.
For analysis, we track binding-induced changes like
\begin{eqnarray}
\Delta U & \equiv & U^{\rm A} +  U^{\rm B} - U^{\rm AB} \, , \\
\Delta T_ {\rm KS} & \equiv & T_{\rm KS}^{\rm A} +  
T_{\rm KS}^{\rm B} - T_{\rm KS}^{\rm AB} \, \\
\Delta T_ {\rm c}^{\rm nl} & \equiv & T_{\rm c}^{\rm nl,A} +  
T_{\rm c}^{\rm nl,B} - 
T_{\rm c}^{\rm nl,AB} \, .
\end{eqnarray}
We also track corresponding expressions for binding-induced
changes in, for example, the total nonlocal-correlation term $\Delta E_{\rm c}^{\rm nl}$. 
In our discussion, we call such differences \textit{binding contributions}.\footnote{The wording `binding contribution' is used to describe any component of the molecular binding even if,
for example, $\Delta T_{\rm c}(\vec{r})$ is negative. Similarly, we 
use the word `spatially resolved binding contributions' to describe binding-induced changes in the spatial variation of, for example,
XC energy terms, like $\Delta e_{\rm c}^{\rm nl}(\vec{r})$; Again this
term is used without regards to the sign of the integrated values.}

Computational results for the binding-induced changes in the KS kinetic energy 
$\Delta T_{\rm KS}$ and in the mean-field electrostatic energy
$\Delta U[n]$ often suffice for a characterization of covalent bonds in 
molecules and materials.\cite{EDA12} This is because 
the combination allows us to characterize and understand orbital 
hybridization.\cite{EDA12,BondAnalysisPap,rohrer11p165423} 
However, for noncovalent bonds we have to look further than changes
in $T_{\rm KS}$. One can generally sort chemical bonds from knowledge of 
the average orbital energy.\cite{RahHof16} The average orbital energy 
is a measure that will, in principle, reflect all correlation effects, 
including those that are manifested in  the kinetic energy. 

A formal analysis of the DFT variational scheme\cite{Levy85} shows that 
\begin{equation}
T_{\rm c}[n]=-E_c[n]+\left[ \frac{\partial E_c[n_{\alpha}]}{\partial \alpha} \right]_{\alpha=1}
\label{eq:LeviVers}
\end{equation}
where $\alpha \equiv 1/\lambda$, Refs.\ \onlinecite{Levy85,Burke97}.
Using the density-scaling analysis, it immediately follows that 
\begin{equation}
T_{\rm c}[n] =E_{\rm c}[n]-E_{{\rm c},\lambda=1}[n]= E_{{\rm xc}}[n]-E_{{\rm xc},\lambda=1}[n] \, ,
\label{eq:scalingextent}
\end{equation}
A similar equation connects $T_{\rm c}^{\rm nl}[n]$ and $E_{\rm c}^{\rm nl}[n]$.
For any given system (solution density $n$), we use numerical differentiation 
to compute $E_{{\rm c},\lambda}[n]$ and $E_{{\rm c},\lambda}^{\rm nl}[n]$ from 
Eq. (\ref{eq:scaleExc}), and
$T_{\rm c}[n]$ and $T_{\rm c}^{\rm nl}[n]$ from Eq.\ (\ref{eq:LeviVers}).

The electron-electron interaction effects in the physical systems are 
now formally available for computation (in the approximations that define $E_{\rm xc}[n]$). 
In particular, we can study the electrodynamical coupling among plasmons\cite{jerry65,ma,ra,hybesc14}
at $\lambda=1$ since $E_{{\rm c},\lambda=1}^{\rm nl}[n]$ is available via
Eq.\ (\ref{eq:scalingextent}). This value $E_{{\rm c},\lambda=1}^{\rm nl}[n]$ is the nonlocal-correlation 
part of $E_{{\rm xc},\lambda=1}[n]$ which, by definition, is given by a contour integral of 
the response $\chi_{\lambda=1}$ evaluated at full electron-electron interaction strength,
Eqs.\ (\ref{eq:lambdahole}) and (\ref{eq:ExcVar}). 

We note in passing that $E_{{\rm c},\lambda=1}^{\rm nl}[n]$ is also the nonlocal-correlation
part of the electron-electron interaction expectation value
\begin{equation}
    E_{{\rm c},\lambda=1}^{\rm nl}[n] = \langle \hat{V} \rangle_{\rm c}^{\rm nl} 
    \equiv \langle \hat{V} \rangle - U[n] - E_{\rm x}[n] - E_{{\rm c},\lambda=1}^{\rm LDA}[n]
    \, .
    \label{eq:FullEcnlEquiv}
\end{equation}
Since the XC energy functional is defined
$E_{\rm xc}=\langle \hat{V}+\hat{T}\rangle -T_{\rm KS} - U[n]$ we can use Eq.\ (\ref{eq:scalingextent})
for the formal identification
\begin{equation}
    E_{{\rm xc},\lambda=1}[n] = \langle \hat{V} \rangle - U[n] \, .
    \label{eq:FullExcEquiv}
\end{equation}
The formal equivalence Eq.\ (\ref{eq:FullEcnlEquiv}) follows by subtracting the 
LDA and gradient-corrected exchange components.

\begin{figure}
\includegraphics[width=0.45\textwidth]{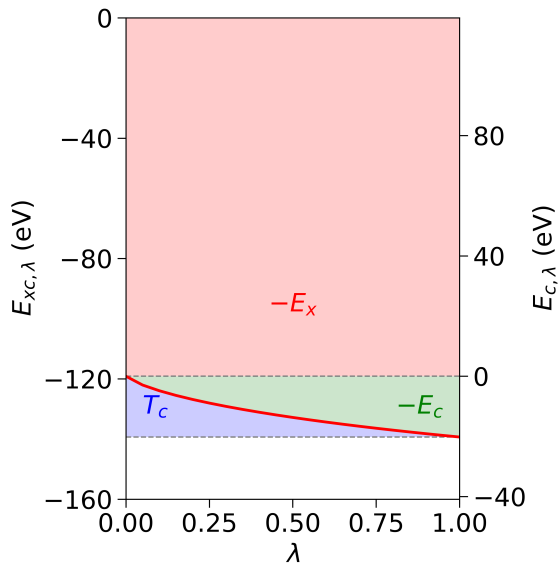}
\caption{Coupling-constant scaling of the vdW-DF-cx exchange and correlation
contributions to the total energy of the N$_2$ molecule. The solid curve
(upper dashed line) shows variation in total XC term
$E_{{\rm xc},\lambda}[n]$ (lack of variation in exchange
term $E_{{\rm x},\lambda}[n]$) for the N$_2$ solution 
electron density $n(\mathbf{r})$. The area of the green 
(red) regions is minus the total correlation (exchange) energy, 
while the area of the blue region is the so-called kinetic-correlation
energy, that is, the kinetic-energy part of correlation, $T_{\rm c}[n]$.}
\label{fig:scalingMol}
\end{figure}

Figure \ref{fig:scalingMol} shows (computed results for) the coupling constant scaling 
for the XC contribution (solid red curve) to the total energy of the N$_2$ molecules. 
The specific scaling results are here provided for vdW-DF-cx (using the formal 
derivation of the scaling for $E_{\rm c}^{\rm nl}$ detailed in the following 
subsection): However, the behavior is generic and thus similar to what has previously been reported and discussed for PBE.\cite{Burke97,Ernzerhof97}

We note that the exchange and correlation components, $E_{\rm x}$ and $E_{\rm c}$, used for DFT calculations in the KS scheme, are integrals of the 
indicated $\lambda$ variations. The 
exchange value traces a horizontal line (dotted curve separating red and green areas) in 
Fig.\ \ref{fig:scalingMol}.  In contrast, the correlation begins at zero but changes to a significant magnitude at $\lambda = 1$. It is straightforward to verify\cite{Burke97} that the area of the green region is minus the functional
approximation for $E_{\rm c}$. Importantly, we can immediately extract the corresponding kinetic-correlation energy  $T_{\rm c}$ using Eq.\ (\ref{eq:scalingextent}), that is, as the area of the blue region below the $E_{{\rm xc},\lambda}$ variation but above the value of the 
$\lambda \to 1$ limit.

\subsection{Coupling-constant scaling and kinetic-correlation energy
in vdW-DF-cx}

\begin{figure}
\includegraphics[width=0.9\columnwidth]{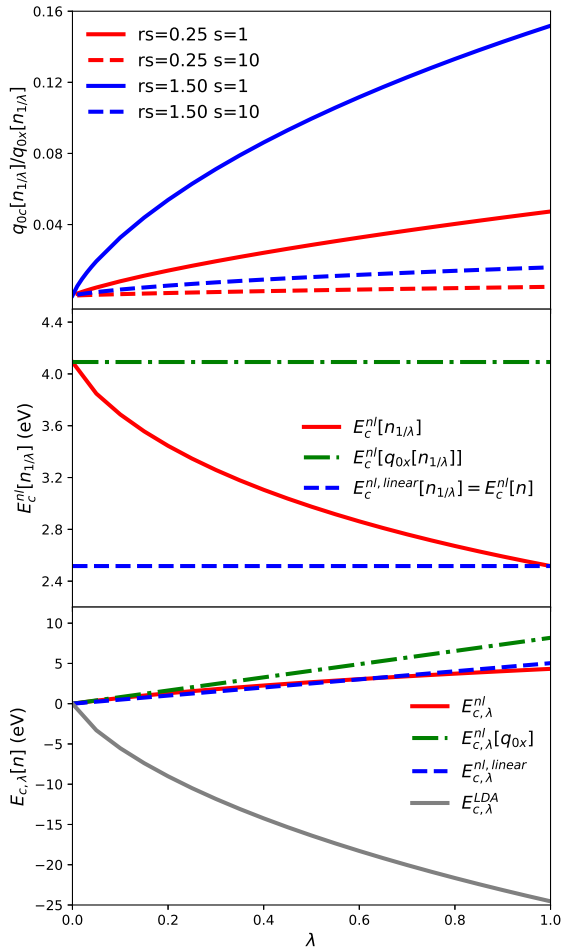}
\caption{Coupling-constant scaling of the nonlocal correlation energy
functional $E_{{\rm c},\lambda}^{\rm nl}[n]$ at various density conditions
typical of the N$_2$ molecule. The binding is thus analyzed in terms of
values for the inverse length scales, $q_0 = q_{0x}+q_{0c}$, that specify
the vdW-DF plasmon model.  \textit{The top panel\/} shows the scaling in the 
ratio $q_{0c}/q_{0x}$ at conditions typical for the inter-atom binding region 
of the N$_2$ molecules (for the nonbinding regions outside
each atom), red (blue) solid curves. The set of dashed curves shows 
a corresponding characterization for density tails.
\textit{The middle panel\/} shows the resulting scaling of 
$E_{\rm c}^{\rm nl}[n_{1/\lambda}]$ (solid curve) with limits 
discussed in the text. \textit{The bottom panel\/} contrasts the resulting
scaling in  $E_{{\rm c},\lambda}^{\rm nl}[n]$ (solid red curve)
against that of LDA correlation (solid grey curve), and as obtained
in two approximations.
\label{fig:ScaleEcnl}
}
\end{figure}

To compute the kinetic-energy component of vdW binding, we need only consider the density 
scaling for the correlation parts, namely $E_c=E_c^{\rm LDA}+E_c^{\rm nl}$. Moreover,
the coupling constant scaling for the LDA part, $E_c^{\rm LDA}$, has previously been discussed, as it is part of the GGA characterization.\cite{Levy85,LYP85,Levy91,Gorling93,Levy95chapter,Levy96,Perdew96,Ernzerhof97,Burke97} 

To explore the coupling-constant scaling of 
$E_{\rm c}^{\rm nl}$ we first summarize the vdW-DF
formulation of this nonlocal-correlation energy.
Any semilocal XC density functional can be characterized by a local 
energy-per-particle density
\begin{equation}
E_{\rm xc}[n]=\int_{\vec{r}} \, 
n(\vec{r})\epsilon_{\rm xc}[n](\vec{r}) \, ,
\end{equation}
where $\epsilon_{\rm xc}[n](\vec{r})$ is a function of just the
local density $n(\vec{r})$ and the scaled density gradient $s(\vec{r})$.
We further split $\epsilon_{\rm xc}[n](\vec{r})$ into 
exchange and correlation components, $\epsilon_{\rm x}[n](\vec{r})$ and 
$\epsilon_{\rm c}[n](\vec{r})$.  The local variation in the 
inverse length scale $q_0$ for the plasmon-pole description can 
then be expressed \cite{Dion,thonhauser}
\begin{eqnarray}
q_0(\vec{r}) & = & q_{0x}(\vec{r}) +  q_{0c}(\vec{r}) \, , \\
q_{0x(c)}(\vec{r}) & = & \frac{\ve_{\rm x(c)}^{\rm in}(\vec{r})}
{\ve_x^{\rm LDA}(\vec{r})}\, k_F(\vec{r}) \, .
\label{eq:q0xcforms}
\end{eqnarray}
Here $k_F=(3\pi^2 n)^{1/3}$ denotes the local value of the Fermi wave-vector
and $\ve_x^{\rm LDA}=-3k_F/4\pi$ is the energy-per-particle density in LDA exchange.
The nonlocal-correlation energy is computed\cite{Dion}
\begin{equation}
E_{\rm c}^{\rm nl}=\frac{1}{2}\int_{\vec{r}}\int_{\vec{r'}} n(\vec{r})\phi(\vec{r},\vec{r}')n(\vec{r}').
\label{eq:Ecnlexpr}
\end{equation}
using a universal-kernel formulation,\cite{Dion,dionerratum,thonhauser,roso09} 
\begin{eqnarray}
 \phi(\vec{r},\vec{r}') & = & \Phi_0(d,d') \,, \\
d(\vec{r},\vec{r'}) & = & |\vec{r}-\vec{r}'| q_0(\vec{r}) \,,\\
d'(\vec{r},\vec{r'}) & = & |\vec{r}-\vec{r}'| q_0(\vec{r'}) \, . 
\end{eqnarray}
The universal kernel $\Phi_0(d,d')$ is tabulated and permits 
an efficient numerical evaluation through fast-Fourier transforms.\cite{roso09} 
The generalization to scaling in spin-polarized cases is also completely 
specified, since it amounts to a simple rescaling of the inverse length scale $q_0$, 
Ref.\ \onlinecite{Thonhauser_2015:spin_signature}.

For scaling analysis (and coding), it is convenient to introduce 
$\tilde{\vec{r}}\equiv \vec{r}/\lambda$ as a short 
hand for the coordinate scaling and to represent the 
density variation in terms of $r_s(\vec{r})= 
(3/4\pi n(\vec{r}))^{1/3}$. The local values of the
scaled density gradient are $s(\vec{r})= |\nabla n|/(2k_F(\vec{r}) n(\vec{r}))$.
The density scaling $n(\vec{r}) \to n_{1/\lambda}(\vec{r})=
n(\tilde{\vec{r}})/\lambda^3$ leaves $s$ unchanged and the
effect amounts to computing the changes in Fermi vector and in the LDA 
correlation components. This is done in terms
of the corresponding scaling $r_s(\vec{r}) \to \lambda r_s(\tilde{\vec{r}})$.

Using $F_x(s)$ to denote the  exchange-enhancement factor of $E_{\rm xc}^{\rm in}$, the 
overall scaling of the inverse length scale can be expressed
$q_0[n_{1/\lambda}(\mathbf{r})] = q_0^\lambda(\tilde{\vec{r}})$, where 
\begin{equation}
q_0^{\lambda}(\tilde{\vec{r}})
= F_x(s(\tilde{\vec{r}})) \, \frac{k_F(\tilde{\vec{r}})}{\lambda} 
- \frac{4\pi}{3} \ve_{\rm c}^{\rm LDA} 
(\lambda r_s(\tilde{\vec{r}})) \, .
\end{equation}
Introducing also 
\begin{eqnarray}
\tilde{d}_\lambda & \equiv & 
q_0^{\lambda}(\tilde{\vec{r}}) \, (\lambda |\tilde{\vec{r}}-\tilde{\vec{r'}}|) \, , 
\label{eq:dscale1}\\
\tilde{d'}_\lambda & \equiv & 
q_0^{\lambda}(\tilde{\vec{r'}}) \, (\lambda |\tilde{\vec{r}}-\tilde{\vec{r'}}|) \, , 
\label{eq:dscale2}
\end{eqnarray}
we can compute
\begin{equation}
	E_{\rm c}^{\rm nl}[n_{1/\lambda}] = \frac{1}{2} \int_{\vec{\tilde{r}}}\int_{\vec{\tilde{r}'}}\,
 n(\tilde{\mathbf{r}}) \,
 \Phi_0( \tilde{d}_{\lambda},  \tilde{d'}_\lambda) \,
  n(\tilde{\mathbf{r'}}) \, .
 \label{eq:ecnl_scale}
\end{equation}

We complete the scaling analysis of $E_{\rm c}^{\rm nl}$ via Eq.\ (\ref{eq:scaleExc}) and of $e_{\rm c}^{\rm nl}(\vec{r})$ via Eq.\ (\ref{eq:scaleEcdensity}). Specifically, we express the scaling of the nonlocal-correlation energy density 
\begin{equation}
    e_{{\rm c},\lambda}^{\rm nl}(\vec{r}) = 
    \frac{d}{d\lambda} \left\{ \lambda^2 e_{\rm c}^{\rm nl}[\lambda r_s(\tilde{\vec{r}})] \right\} \, .
    \label{eq:ecnlscalelambda}
\end{equation}
For the numerical evaluation we adopted the scheme proposed by Rom\'an-P\'erez and Soler~\cite{roso09} (as implemented in \textsc{quantum-espresso}) to calculate $e_{\rm c}^{\rm nl}[\lambda r_s(\mathbf{\tilde{r}})]$; the 
calculation is similar to the calculation of $e_{\rm c}^{\rm nl}(\vec{r})$ in Ref.~\onlinecite{callsen12p085439}. We note that 
$e_{{\rm c},\lambda=1}(\vec{r})$ provides a spatial 
mapping of all nonlocal-correlation effects that exist in
the fully interacting system, as a direct implication of 
Eq.\ (\ref{eq:scalingextent}).

Moreover, as part of this $e_{{\rm c},\lambda=1}(\vec{r})$ characterization,
we can now compute the spatially resolved kinetic-correlation energy
$t_{\rm c}(\vec{r})$ and nonlocal-kinetic-correlation energy $t_{\rm c}^{\rm nl}(\vec{r})$. 
First, we simply add the
known\cite{Levy85,Levy91,Gorling93,Levy95chapter} coupling constant scaling 
of the LDA correlation-energy density, $e_{{\rm c},\lambda}^{\rm LDA}$, 
entering in Eq.\ (\ref{eq:LDAcorscale}): 
\begin{equation}
    e_{{\rm c},\lambda}(\vec{r}) = e_{{\rm c},\lambda}^{\rm LDA}(\vec{r}) + 
    e_{{\rm c},\lambda}^{\rm nl}(\vec{r}) 
    \, .
    \label{eq:ecscalelambda}
\end{equation}
Next, we adapt Eq.\ (\ref{eq:LeviVers}) to descriptions of energy densities
\begin{eqnarray}
    t_{\rm c}(\vec{r}) & = & - e_{\rm c}(\vec{r}) 
    + \left[ 
	\frac{\partial e_{\rm c}[r_s(\mathbf{\tilde{r}})/\alpha]}{\partial \alpha} 
\right]_{\alpha=1} \, ,
    \label{eq:tcdef} \\
    t_{\rm c}^{\rm nl}(\vec{r}) & = & - e_{\rm c}^{\rm nl}(\vec{r}) +
    \left[
   \frac{\partial e_{\rm c}^{\rm nl}[r_s(\mathbf{\tilde{r}})/\alpha]}{\partial \alpha} 
\right]_{\alpha=1} \, ,
    \label{eq:tcnldef} 
\end{eqnarray}
and evaluate the derivatives numerically.

\begin{figure*}
\includegraphics[width=0.75\textwidth]{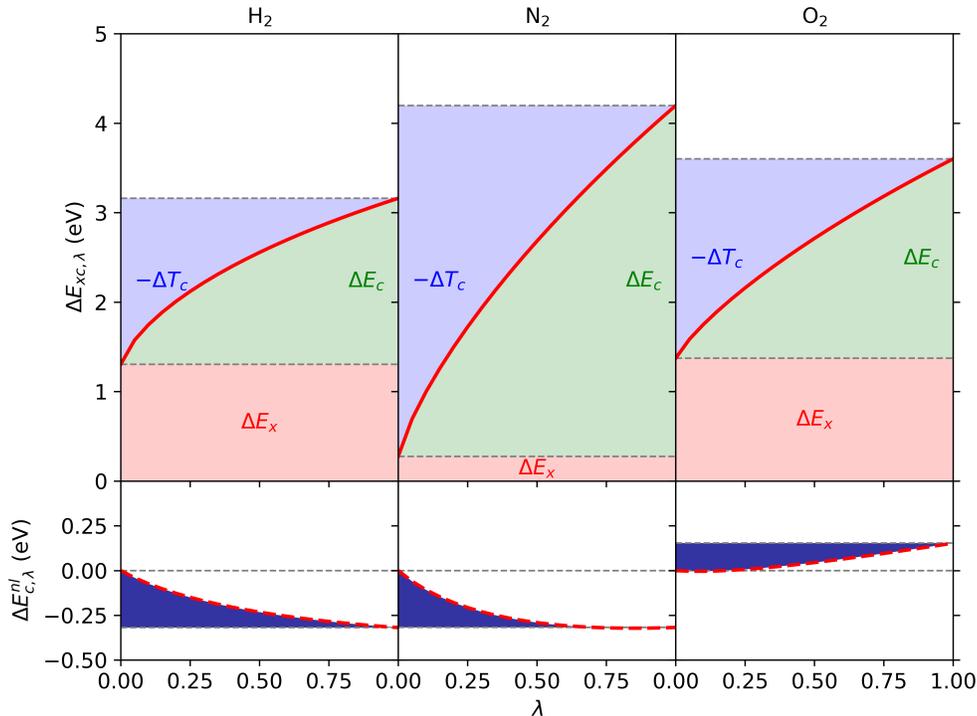}
\caption{Coupling-constant scaling for the vdW-DF-cx exchange and correlation
contributions to the atomization energy of the H$_2$, N$_2$ and O$_2$ 
molecules. The bottom panel shows the scaling in the nonlocal-correlation 
contribution to the molecular bindings. The dark blue area is here a measure 
of the magnitude (108, 67, and -99 meV for H$_2$, N$_2$, and O$_2$) 
of $\Delta T_{\rm c}^{\rm nl}$, i.e., the binding contribution arising in the 
nonlocal part of the kinetic-correlation energy, Eq.\ (\ref{eq:scalingextent}).
}
\label{fig:scalingMolAE}
\end{figure*}

Figure \ref{fig:ScaleEcnl} documents the coupling-constant scaling of 
$E_{{\rm c},\lambda}^{\rm nl}$ and explains an approximately linear 
variation.
The top panel shows the coupling-constant scaling for 
the ratio $q_{0c}/q_{0x}$ for typical contributions to the binding
of an N$_2$ molecule. Specifically, starting from the known 
$q_{0x}(n(\vec{r}))$ and $q_{0c}(n(\vec{r}))$ values, the panel traces 
the variation in the XC components of $q_0(n_{1/\lambda}(\vec{r}))$
for conditions that roughly correspond to the binding region of N$_2$ 
(red curves) and to electron density tails of atoms and molecules
(blue curves).  

The middle panel of Fig.\ \ref{fig:ScaleEcnl} shows the coupling-constant
scaling of $E_{\rm c}^{\rm nl}[n_{1/\lambda}]$ (solid curve) for the density of
the N$_2$ molecule.  We note that the scaling in $q_{0x}$ is exactly offset by 
the $\lambda$ scaling of coordinates in Eqs.\ (\ref{eq:dscale1}) and (\ref{eq:dscale2}). 
Thus if we assume that the scaling of 
$q_0 = q_{0x}+q_{0c}$ is set by the scaling in $q_{0x}$, there would be 
no $\lambda$-dependence in the $E_{\rm c}^{\rm nl}$ kernel arguments, 
$d=q_0(\vec{r})|\vec{r}-\vec{r'}|$ and $d'=q_0(\vec{r'})|\vec{r}-\vec{r'}|$.
In this type of approximations there is then no scaling in the corresponding 
approximations for $E_{\rm c}^{\rm nl}[n_{1/\lambda}]$.

The middle panel furthermore shows two potentially relevant such approximations  
motivated by the analysis of the typical variations in the $q_{0c}/q_{0x}$ 
ratio. The first assumes that we can ignore the influence of the correlation part
$q_{0c}$ completely (giving the green dashed-dotted line); the second assumes that 
the ratio $q_{0c}/q_{0x}$ can at any given point $\vec{r}$ be taken as fixed at 
the $\lambda=1$ value (giving the blue dashed curve).  The second choice 
effectively amounts to simply setting $E_{\rm c}^{\rm nl,linear}[n_{1/\lambda}] 
\equiv E_{\rm c}^{\rm nl}[n]$. Neither of them is a good description for 
Eq.\ (\ref{eq:ecnl_scale}). On the other hand, we add a $\lambda^2$ weight
on $E_{\rm c}^{\rm nl}[n_{1/\lambda}]$ when computing $E_{{\rm xc},\lambda}$, 
Eq.\ (\ref{eq:scaleExc}). We label the second approximation as `linear' 
since it leads to $E_{{\rm c},\lambda}^{\rm nl}\approx \lambda E_{\rm c}^{\rm nl}[n]$
and this is sometimes an acceptable approximation.

The bottom panel of Fig.\ \ref{fig:ScaleEcnl} contrasts the resulting scaling
of the ACF integrand for the nonlocal correlation contribution $E_{{\rm c},\lambda}^{\rm nl}[n]$ 
against that of the LDA correlation $E_{\rm c}^{\rm LDA}[n]$. The panel also shows (green and 
blue dashed curves) the scaling that results by inserting either of the approximations 
discussed in the middle panel into Eq.\ (\ref{eq:scaleExc}). Interestingly, 
the scaling relevant for the N$_2$-molecule total energy is found well approximated by using
$E_{\rm c}^{\rm nl,linear}[n_{1/\lambda}]\equiv E_{\rm c}^{\rm nl}[n]$.
The nonlocal-correlation part of the kinetic energy is just minus
the total nonlocal-correlation binding contribution $\Delta E_{\rm c}^{\rm nl}$
in such special cases. 

\begin{figure*}
\includegraphics[width=0.65\textwidth]{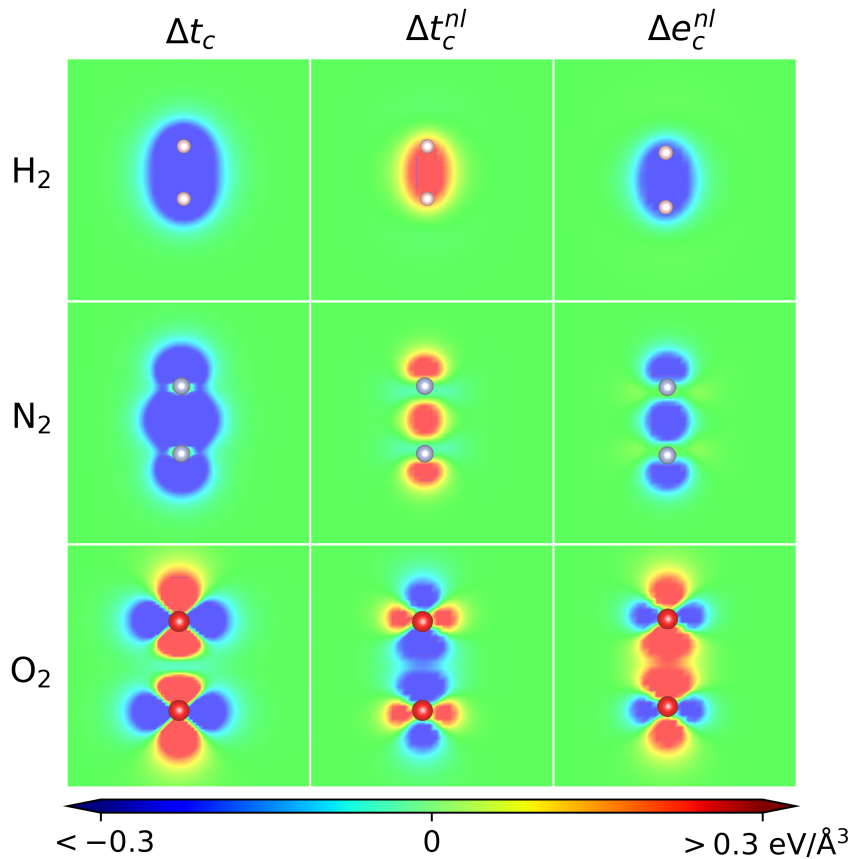} 
\caption{\label{fig:SpaceVarCov_ts} 
Spatially resolved kinetic-correlation and nonlocal-correlation energy 
contributions to the atomization energy for H$_2$, N$_2$ and O$_2$ molecules.
The color map is the energy density in eV/\AA$^3$.  
The first and second columns show maps of the full kinetic-correlation 
binding contribution $\Delta t_{\rm c}(\vec{r})$ and of the nonlocal-correlation kinetic energy contribution $\Delta t_{\rm c}^{\rm nl}(\vec{r})$. The third column shows the binding-energy 
contribution $\Delta e_{\rm c}^{\rm nl}(\vec{r})$ that directly
reflects $E_{\rm c}^{\rm nl}$.
}
\end{figure*}

\section{Computational details }

We focus our discussion and mapping of many-body physics effects in results obtained
using the vdW-DF-cx version.\cite{behy14} In vdW-DF-cx, the total exchange component in Eq.\ (\ref{eq:fullvdWDF}) is picked  so that binding contributions from $\delta E_{\rm x}^0$ can  generally be ignored.\cite{behy14,hybesc14} 

The vdW-DF-cx version performs well, on par with or better than
PBE, for characterizations of many bulk, surface, and interface properties.\cite{bearcoleluscthhy14,Berland_2015:van_waals,RanPRB16,BrownAltvPRB16,AmbSil16,Gharaee2017,PetArn17,LonPopDes17,Olsson17,WaZhKe17,WaEsZe17} The vdW-DF-cx version 
has proven itself useful also in the description of binding and 
function of layered materials, at surfaces, and of
molecules.\cite{torbjorn14,ErhHylLin15,SadSanLam15,RaiPRB16,LinErh16,FriSerSol16,JavAkh16,Arnau16,HeBeBr16,MehDorZhu16,MehFreYan16,LofErh16,KuiHanLin16,CecKleArn16,ZhoWan17,BorArn17,BriYnd17,KeSpMi17,BorSch17,DFcx02017}

Our calculations are based on the plane-wave \texttt{Quantum Espresso}
package,\cite{QE} which already has the consistent exchange vdW-DF-cx
version \cite{behy14} as well as the rigorous spin extension
of the vdW-DF method.\cite{thonhauser} Core electrons are
represented by Troullier-Martins type normal-conserving
pseudo potentials using a  80~Ry wavefunction cutoff.

This paper also introduces a post-processing ACF-analysis
code, termed \textsc{ppACF}, which tracks the 
system-specific coupling constant variation in 
$E_{\rm xc}[n]$ (for standard GGA and vdW-DF versions).
The code adapts the post-processing components of the 
\textsc{quantum espresso} package,\cite{QE} into which \textsc{ppACF} 
will also be released.

The \textsc{ppACF} code takes as input the set of \textsc{quantum-espresso}
solution files (available after completion of the DFT calculations).
It outputs the coupling-constant scaling analysis and the
spatial variation in the kinetic-correlation energy density. 
For convenience, it also outputs the spatial variation in 
the set of XC components.

Our numerical analysis is based on comparing binding-energy contributions 
for the various components of the total DFT description. To discuss the 
binding `AB' of fragments  `A' and `B',  the \textsc{ppACF} code outputs 
the spatial variation in all XC components. The code furthermore uses the coupling-constant scaling in the spatially resolved correlation terms, $e_{\rm c}$ and $e_{\rm c}^{\rm nl}$, to numerically determine and output (for any given fragment and for the combined system) the spatial variation in $e_{{\rm c},\lambda=1}$ and  $e_{{\rm c},\lambda=1}^{\rm nl}$ as well as in $t_{{\rm c},\lambda=1}$ and  $t_{{\rm c},\lambda=1}^{\rm nl}$. We then obtain, for the (spatially-resolved) binding-energy contributions 
\begin{eqnarray}
\Delta e_{{\rm c},\lambda=1} & = &
e_{{\rm c},\lambda=1}^{\rm A} + e_{{\rm c},\lambda=1}^{\rm B} - e_{{\rm c},\lambda=1}^{\rm AB} \, , \\
\Delta e_{{\rm c},\lambda=1}^{\rm nl} & = &
e_{{\rm c},\lambda=1}^{\rm A,nl} + e_{{\rm c},\lambda=1}^{\rm B,nl} - e_{{\rm c},\lambda=1}^{\rm AB,nl} \, , \\
\Delta t_{{\rm c},\lambda=1} & = &
t_{{\rm c},\lambda=1}^{\rm A} + t_{{\rm c},\lambda=1}^{\rm B} - t_{{\rm c},\lambda=1}^{\rm AB} \, , \\
\Delta t_{{\rm c},\lambda=1}^{\rm nl} & = &
t_{{\rm c},\lambda=1}^{\rm A,nl} + t_{{\rm c},\lambda=1}^{\rm B,nl} - t_{{\rm c},\lambda=1}^{\rm AB,nl} \, ,
\end{eqnarray}
from simple numerical subtractions.  

For completeness, the \textsc{ppACF} code outputs the spatial variation 
in KS kinetic energy $t_{\rm KS}(\vec{r})$ and in a spatially resolved
measure of the full kinetic energy
\begin{equation}
t_{\rm tot}(\vec{r}) \equiv t_{\rm KS}(\vec{r})+t_{\rm c}(\vec{r}) \, .
\label{eq:ttotvardef}
\end{equation}
Again by numerical subtractions we can then define spatially-resolved 
kinetic binding energy contributions
\begin{eqnarray}
\Delta t_{\rm KS}(\vec{r}) & = & 
t_{\rm KS}^{\rm A}(\vec{r}) + t_{\rm KS}^{\rm B}(\vec{r})
-  t_{\rm KS}^{\rm AB}(\vec{r}) \, , 
\label{eq:dtksdef}\\
\Delta t_{\rm tot}(\vec{r}) & = & 
t_{\rm tot}^{\rm A}(\vec{r}) + t_{\rm tot}^{\rm B}(\vec{r})
-  t_{\rm tot}^{\rm AB}(\vec{r}) \, .
\label{eq:dktotdef}
\end{eqnarray}
A mapping of the total kinetic energy
binding contributions Eq.\ (\ref{eq:dktotdef}) will, in principle, 
always change if we base the $t_{\rm tot}$ definition, 
Eq.\ (\ref{eq:ttotvardef}), on $\tilde{t}_{\rm occ}$ instead of on
$\tilde{t}_{\rm KS}$, using Eq.\ (\ref{eq:tkschanges}). This is true 
even if the integral values $\Delta T_{\rm KS}$ and $\Delta T_{\rm tot}$ 
remain the same. Qualitative differences in the resulting total-kinetic-energy 
mappings are visible for covalent bonding, but not for the cases of 
noncovalent inter-molecular interactions that we have investigated.

The set of top panels of Fig.\ \ref{fig:scalingMolAE} compares the coupling 
constant variation in the contributions $\Delta E_{{\rm x},\lambda}$ and 
$\Delta E_{{\rm c},\lambda}$ to the H$_2$, N$_2$, and O$_2$ atomization energies,
as computed in vdW-DF-cx.  The scaling and the
total kinetic-correlation energy contributions vary significantly between 
these traditional molecular binding examples. The total kinetic-correlation
energy contribution to binding $\Delta T_{\rm c}$ is given by the light blue area 
under the scaling curve. The value of $\Delta T_{c}$ is dominated by the part that 
originates from the LDA correlation energy.

The set of bottom panels focus on the coupling-constant scaling of the nonlocal-correlation
contribution to the molecular cohesion.  The coupling-constant variation in 
$E_{{\rm c},\lambda}^{\rm nl}$ can be either upward or downward concave because it is
only a part of the kinetic-correlation energy. The upwards and downwards concave behavior
corresponds to positive and negative values of $\Delta T_{\rm c}^{\rm nl}$ binding 
contributions, respectively. The dark-blue areas indicate the magnitude of this 
binding contribution. The supplementary materials includes a broad listing and comparisons
of molecular-binding contributions $\Delta E_{\rm c}^{\rm nl}$,
$\Delta T_{\rm c}$ and $\Delta T_{\rm c}^{\rm nl}$. The comparison also lists
KS binding contributions $\Delta T_{\rm KS}$, making it clear that the kinetic-correlation
energy can only play a significant role in the case of inter-molecular binding.

In the case of binding in the H$_2$, N$_2$, and O$_2$ molecules, Fig.\ \ref{fig:scalingMolAE}, 
we observe that the nonlocal-correlation
contribution to binding is offset by a contribution to the 
nonlocal part of the correlation-kinetic energy. As shown in 
the supplementary materials, the same is true for 
many intra-molecular bonds,  Tables S.I and S.II, and for all
investigated inter-molecular interaction cases, Table S.III.

The binding in the total correlation term, $\Delta E_{\rm c}$, will be offset by a negative kinetic-correlation energy contribution $\Delta T_{\rm c}$, as suggested by the virial theorem.
However, this need not hold generally for the nonlocal part of the kinetic-correlation energy
contribution for intra-molecular binding, as further documented in the supplementary
materials, Tables S.I and S.II. On the other hand, the compensation can be expected when 
the nonlocal part of the correlation-kinetic energy is a significant 
component, such as in most inter-molecular interactions.
\section{Kinetic-energy mappings of molecular binding}

We analyze and discuss the nature of binding both in H$_2$, N$_2$, and 
O$_2$ molecules (having traditional chemical bonds) and in 
non-covalently bonded systems (where, in contrast, there is no pronounced orbital hybridization).

\subsection{Intra-molecular interactions}

Figure \ref{fig:SpaceVarCov_ts} shows that the kinetic-correlation energy is 
important in characterizations of intra-molecular binding. The figure details
the spatial variation in the kinetic-correlation binding 
energy contribution $\Delta t_{\rm c}$, in the nonlocal-correlation-kinetic
energy contribution  $\Delta t_{\rm c}^{\rm nl}$, and in the  
vdW-DF-cx nonlocal correlation energy binding contribution 
$\Delta e_{\rm c}^{\rm nl}$ for the H$_2$, N$_2$, and O$_2$ molecules.

We note that the magnitude of the variation in $\Delta t_{\rm c}$  is 
about an order-of-magnitude smaller than the KS kinetic-energy binding contribution $\Delta t_{\rm KS}(\vec{r})$ (not shown) for these
covalently bonded systems. Nevertheless, there is clear structure
in both $\Delta t_{\rm c}^{\rm nl}(\vec{r})$ and $\Delta e_{\rm c}^{\rm nl}(\vec{r})$ and a directed nature or signature of vdW interactions even 
in these strongly bonded dimer molecules.

Supplementary materials Tables S.I 
and S.II, supported by Fig.\ S1, provides a broader analysis of such 
intra-molecular bindings. This is done both for the set of molecules
for which there already exists a PBE-based coupling-constant 
analysis,\cite{Burke97,Ernzerhof97} and for the G2-1 benchmark set 
of molecular atomization energies. Most of these systems are
covalently bonded, meaning that orbital hybridization plays the decisive role. 
However, there are also some G2-1 cases, for example, alkali dimers, where the 
nonlocal-correlation energy and the nonlocal part of kinetic-correlation 
energy are important. Here, in the main text we concentrate on 
characterizing the binding in H$_2$, N$_2$, and O$_2$.

The O$_2$ kinetic-correlation energy binding contribution $\Delta t_{\rm c}$
deserves a special discussion. The first O$_2$ panel of Fig.\ \ref{fig:SpaceVarCov_ts} 
shows the variation of $\Delta t_{\rm c}(\vec{r})$ in a plane that contains 
the binding axis in the dimer. This plot has areas of opposite signs and implies a compensation. However, the overall 
kinetic-correlation energy contribution is still negative, $\Delta T_{\rm c} <0$, because the 
negative regions, away from the axis, have greater weight as we perform the spatial integration.
The total, negative kinetic-correlation energy binding contribution
is given by the light blue area shown in the left column in Fig.\ \ref{fig:scalingMolAE}. 

The second column of Fig.\ \ref{fig:SpaceVarCov_ts} shows the spatial 
variation in the nonlocal-correlation part of the kinetic-energy binding contribution, 
$\Delta t_{\rm c}^{\rm nl}(\vec{r})$, for the three molecules.  The variation in 
$\Delta t_{\rm c}(\vec{r})$ (first column) is generally 
dominated by the LDA contribution but adjusted by the variation in $\Delta t_{\rm c}^{\rm nl}(\vec{r})$. 
The integrated binding contribution from the nonlocal part of the kinetic-correlation
energy $\Delta T_{\rm c}^{\rm nl}$ can be both negative or positive (as exemplified 
by the H$_2$ and N$_2$ cases). In such positive-$\Delta T_{\rm c}^{\rm nl}$ cases, the
binding contribution from $E_{\rm c}^{\rm nl}$ is negative, i.e., the nonlocal-correlation
energy is actually causing a repulsion in these intra-molecular binding cases.
The supplementary materials Tables S.I and S.II provide a broader overview of the 
variation in nonlocal-correlation energy and kinetic-correlation energy effects
that we document for intra-molecular binding. 

The third column of Fig.\ \ref{fig:SpaceVarCov_ts} shows a mapping of 
the nonlocal-correlation energy contribution to binding,  $\Delta e_{\rm c}^{\rm nl}(\vec{r})$,
allowing a contrast with the variation documented for $\Delta t_{\rm c}(\vec{r})$
and $\Delta t_{\rm c}^{\rm nl}(\vec{r})$.
We find that $\Delta e_{\rm c}^{\rm nl}(\vec{r})$ and 
$\Delta t_{\rm c}^{\rm nl}(\vec{r})$ are here essentially  negative prints of each others.
It follows that the full ($\lambda=1$)
nonlocal-correlation energy contribution, given by
$\Delta e_{\rm c}^{\rm nl}(\vec{r})- \Delta t_{\rm c}^{\rm nl}(\vec{r})$
remains qualitatively described by the variation in $\Delta e_{\rm c}^{\rm nl}(\vec{r})$ in these cases.

Supplementary materials Tables S.I and S.II show that the nonlocal-correlation
energy binding contribution $\Delta E_{\rm c}^{\rm nl}$
can take either sign for intra-molecular binding. These tables also
show that $\Delta T_{\rm c}^{\rm nl}$ will typically then have the opposite sign. We therefore generally expect the $\Delta e_{\rm c}^{\rm nl}(\vec{r})$
and $-\Delta t_{\rm c}^{\rm nl}(\vec{r})$ contributions to mirror each other, 
as in Fig.\ \ref{fig:SpaceVarCov_ts}. The implication is that $\Delta e_{\rm c}^{\rm nl}(\vec{r})$ provides a qualitatively correct mapping of the vdW interaction in most covalently bonded cases. 

\begin{figure*}
\includegraphics[width=0.95\textwidth]{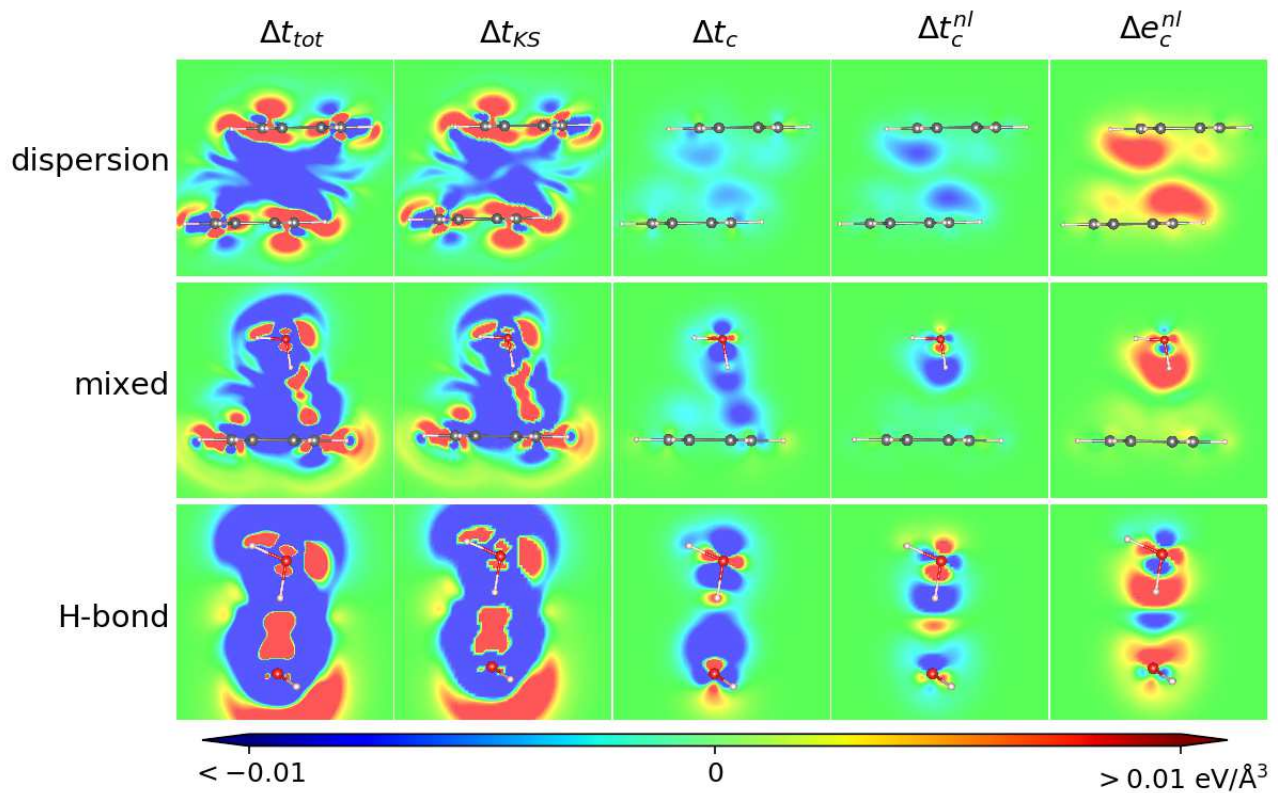}  
\caption{\label{fig:SpaceVarNonCov_ts} Spatial variations in binding
contributions for the benzene-benzene dimer (first row, identified as `dispersion'), 
for the water dimer (last row, `H-bond'),
and for the benzene-water complex (middle row, `mixed'). 
The panels show contours of the binding-energy density, contrasting the
binding contribution arising from the total kinetic energy 
$\Delta t_{\rm tot}$
(first column), the dominant 
KS kinetic-energy component $\Delta t_{\rm KS}$ (second column), 
the total kinetic-correlation energy $\Delta t_{\rm c}$ (third column)
and the non-local correlation component of the kinetic correlation energy 
$\Delta t_{\rm c}^{\rm nl}$ 
(fourth column).
The latter is found to closely track the spatial variations in the binding contribution
from the nonlocal correlation energy variation $\Delta e_c^{\rm nl}$ (last column). 
}
\end{figure*}

Finally, we note that there are exceptions to this general trend, i.e,
cases where $\Delta T_{\rm c}^{\rm nl}$ and $\Delta E_{\rm c}^{\rm nl}$
are both negative. Figures S1 and S2 in the supplementary materials, 
provide additional analysis for one of these cases, namely P$_2$.
In such cases it is in principle necessary to compute both the 
$\Delta e_{\rm c}^{\rm nl}(\vec{r})$ and the 
$\Delta t_{\rm c}^{\rm nl}(\vec{r})$ variation to 
obtain a complete mapping, $\Delta e_{\rm c}^{\rm nl}(\vec{r})-
\Delta t_{\rm c}^{\rm nl}(\vec{r})$, Fig.\ S2. However, even in these
cases it is still so that $\Delta e_{\rm c}^{\rm nl}(\vec{r})$ and the 
$-\Delta t_{\rm c}^{\rm nl}(\vec{r})$ are approximately mirrors of each others. 
That is, even in P$_2$, there is no change in the qualitative 
observations presented above.

\begin{figure}
\includegraphics[width=0.46\textwidth]{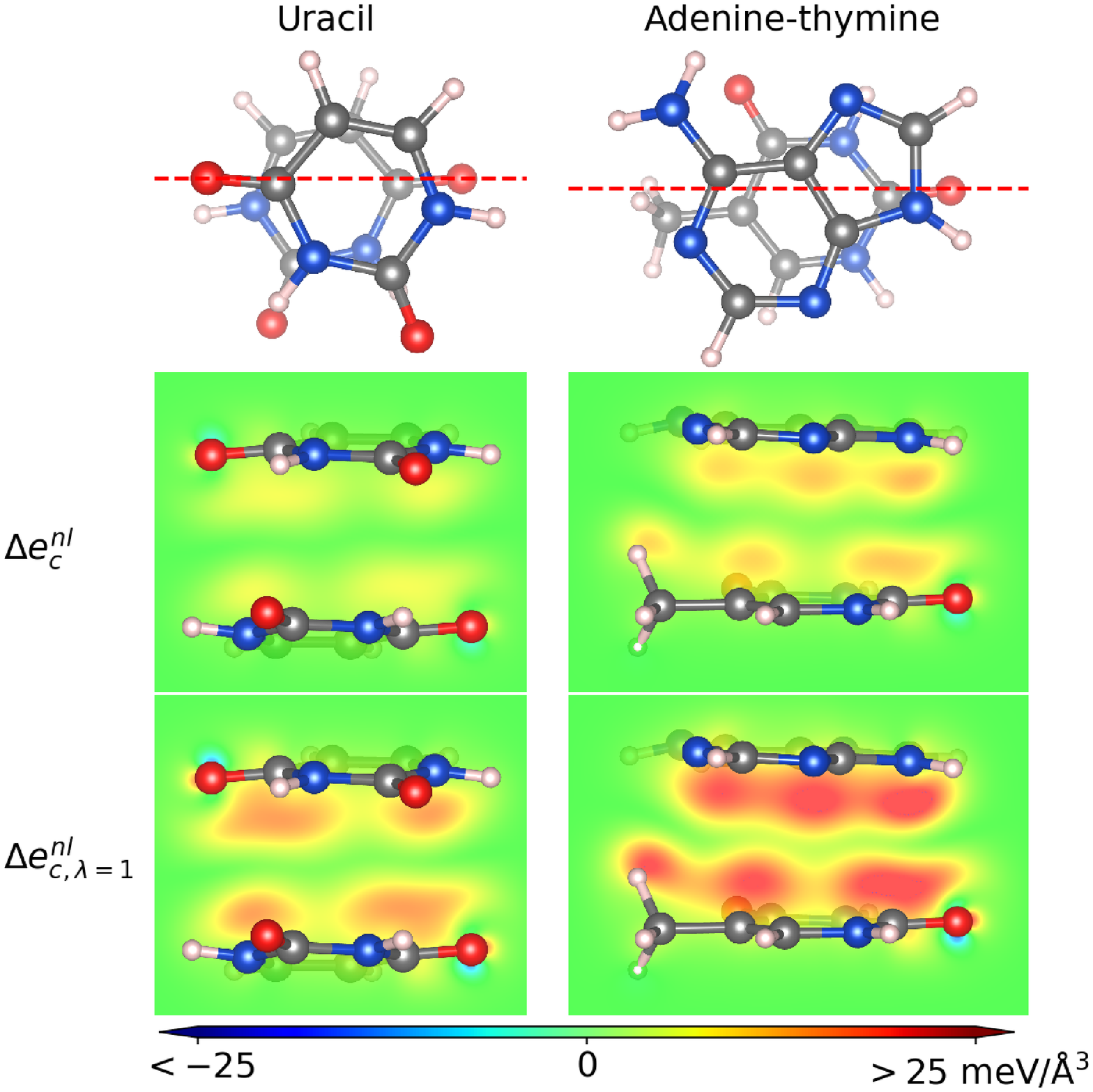}  
\caption{\label{fig:SpaceVarBases} 
Spatial variation in nonlocal-correlation binding contributions to the stacked uracil dimer (left panel) and to the stacking of adenine and thymine bases (right panel). The panels in the first row show top view of the configurations; Configurations are taken at either the CCSD(T) or MP2 level, from Ref.~\onlinecite{S22}. The set of red dashed lines marks the planes plotted 
in which we trace the binding contributions. The panels in the second row 
show the binding contributions $\Delta e_{\rm c}^{\rm nl}$ from the nonlocal correlation energy of the vdW-DF-cx functional. The panels in the third row show the total nonlocal-correlation  binding contribution
$\Delta e_{{\rm c},\lambda=1}^{\rm nl} = \Delta e_{\rm c}^{\rm nl}-\Delta t_{\rm c}^{\rm nl}$, i.e., the spatial variation in the vdW binding of the
physical systems, at full electron-electron interaction.
}
\end{figure}

\subsection{Inter-molecular interactions}

Figure \ref{fig:SpaceVarNonCov_ts} compares kinetic-energy binding contributions
$\Delta t_{\rm tot}(\vec{r})$, $\Delta t_{\rm KS}(\vec{r})$, 
$\Delta t_{\rm c}(\vec{r})$, $\Delta t_{\rm c}^{\rm nl}(\vec{r})$,
in non-covalently bonded systems, i.e., in cases where there 
is no pronounced orbital hybridization.  The top row shows our vdW-DF-cx based
results for a benzene dimer, a case which is expected to have an essentially pure 
vdW (or dispersion) interaction. The bottom row shows results for a water dimer, a case 
that is predominantly hydrogen bonded, while the middle row explores the 
mixed-binding benzene-water case. 

The supplementary materials Table S.III, supported by Fig.\ S3, 
provides a broader characterization of binding in the S22 benchmark set of 
such weakly bonded molecular complexes. In all of theses cases, the nonlocal-correlation
energy contribution $\Delta E_{\rm c}^{\rm nl}$ is positive while
$\Delta T_{\rm c}^{\rm nl}$ is negative. As such, the following discussion 
is generic. 

For reference, the last column of Fig.\ \ref{fig:SpaceVarNonCov_ts}
shows the spatial variation in the nonlocal-correlation energy binding, 
$\Delta e_{\rm c}^{\rm nl}(\vec{r})$. This binding plays a decisive role in 
all cases. It is a core component of our vdW-DF-cx characterization
and it is important for an accurate description of these molecular 
complexes.\cite{behy14} In the case of 
dispersion-bonded systems, the $\Delta e_{\rm c}^{\rm nl}(\vec{r})$ 
contributions are the only sources of cohesion; In the case of the 
water dimer and the benzene-water complexes, there are also significant 
electrostatic components in the inter-molecular interactions.

The first, second, and third columns of Fig.\ \ref{fig:SpaceVarNonCov_ts}
contrast the spatial variation in the binding contributions
from the total kinetic energy, from the  KS kinetic energy, and from the
kinetic-correlation energy. There are no orbital hybridization 
effects in play, but it is important to note that the $E_{\rm c}^{\rm nl}$ binding
also pushes densities and thus orbitals around. Smaller signatures are 
therefore retained in the binding contribution from the KS kinetic energy,
$\Delta t_{\rm KS}(\vec{r})$. 

Contrasting the panels in the first and second column of Fig.\ \ref{fig:SpaceVarNonCov_ts},
we find that the KS kinetic-energy effects are still the major source of the 
variation that we compute for $\Delta t_{\rm tot}(\vec{r})$.
Nevertheless, in the case of dispersion-bonded systems, we find 
that there are also important contributions from the kinetic-correlation 
energy $\Delta t_{\rm c}(\vec{r})$. These contributions, shown in isolation in 
the third column, arise primarily in the inter-molecular region, in areas that
have a sparse\cite{langrethjpcm2009} (but not vanishing) electron density and 
small-to-moderate density gradient. We sometimes refer to these binding
parts as a trough\cite{behy14,hybesc14} but we are then emphasizing the presence 
of important internal surfaces within such 
sparse intermolecular regions.\cite{rydberg03p126402,kleis08p205422,langrethjpcm2009} 

The enhanced binding contributions from internal surfaces reflect the many-electron 
nature of the vdW problem. The amplitudes of collective (plasmon) excitations are 
themselves enhanced in the sparse surface-like intermolecular region and we should
then expect larger contributions to the systematic tracking of the electrodynamical coupling among plasmons.\cite{jerry65,lape77,ma,ra,lavo87,rydberg03p126402,Dion,hybesc14,Berland_2015:van_waals} 
The vdW enhancement can also be interpreted as reflecting image-plane effects at (internal 
or external) surfaces,\cite{zarembakohn1976,zarembakohn1977,harrisnordlander1984,rydberg03p126402,persson2008,lee11p193408,lee12p424213,TaoRappe14} or as multipole response-effects effects when arising outside 
molecules.\cite{becke05p154101,becke07p154108,kleis08p205422,berland10p134705,TaoPerdew14,Tao15,NanovdWScale} 
In any case, the vdW-DF-cx handling of screening\cite{jerry65,ra,Dion,bearcoleluscthhy14,hybesc14}
provides mechanisms to track the expected vdW enhancement in the sparse intermolecular regions, 
at important internal surfaces.\cite{rydberg03p126402,rationalevdwdfBlugel12,behy13,hybesc14}

The fourth column of Fig.\ \ref{fig:SpaceVarNonCov_ts} shows 
the nonlocal part of the kinetic-correlation energy, $\Delta 
t_{\rm c}^{\rm nl}(\vec{r})$. Contrasting the third and fourth columns 
makes it clear that the LDA component of $\Delta t_{\rm c}(\vec{r})$
generally masks the variation in $\Delta t_{\rm c}^{\rm nl}(\vec{r})$.
However, the signatures of the nonlocal kinetic-correlation part 
dominate in the spare-density regions for dispersion-bonded systems. 
The nonlocal part also remains a non-vanishing part of full kinetic-correlation
energy in the hydrogen-bonded and mixed binding cases. 

Figure \ref{fig:SpaceVarBases} shows the computed binding
contributions in the stacked uracil dimer and in the stacking of 
adenine and thymine. The top row shows
the investigated geometries from the S22 benchmark set.\cite{S22}
Supplementary materials Fig.\ S.4 shows the variation in the kinetic-energy binding contributions for these systems.

The panels in the bottom two rows of Figure \ref{fig:SpaceVarBases} contrast the
variation in the $\Delta e_{\rm c}^{\rm nl}$ and 
$\Delta e_{{\rm c},\lambda=1}^{\rm nl}=\Delta e_{\rm c}^{\rm nl}(\vec{r})-\Delta t_{\rm c}^{\rm nl}(\vec{r})$ 
accounts of the nonlocal-correlation binding. 
The contributions are computed for the cuts indicated by the two dashed lines in the top panels. 
We find that including the nonlocal part of the kinetic-correlation energy enhances 
the binding signatures found in the $\Delta e_{\rm c}^{\rm nl}$ variation in our mapping 
of the total nonlocal correlation binding, shown in the pair of lower panels. 
The same is true for the wider set of S22 cases. 
Comparing the fourth and fifth column of Fig.\ \ref{fig:SpaceVarNonCov_ts} 
we see that they are essentially negative prints of each other, i.e., 
naturally leading to an enhancement of signatures in $\Delta e_{\rm c}^{\rm nl}(\vec{r})-\Delta t_{\rm c}^{\rm nl}(\vec{r})$.

For both inter- and intra-molecular interactions the variation,
$\Delta e_{{\rm c},\lambda=1}^{\rm nl}(\vec{r})$ can effectively be 
mapped using either $\Delta e_{\rm c}^{\rm nl}(\vec{r})$
or $-\Delta t_{\rm c}^{\rm nl}(\vec{r})$.

We also find that the $\Delta e_{{\rm c},\lambda=1}^{{\rm nl}}(\vec{r})$ signatures are, 
in effect, channelled into pockets of binding, 
Figs.\  \ref{fig:SpaceVarNonCov_ts} and \ref{fig:SpaceVarBases}.
The dominant contributions are located in the intermolecular (trough) regions 
but concentrated in areas that resemble orbitals. In other words, 
we can, in principle, use a similar form of bond-type characterization for 
a qualitative discussion of the nonlocal-correlation binding, drawing on an
analogy with discussions of chemical bonds. 

Before using this vdW-bond mapping analysis, below, we emphasize that this
binding is much weaker and that the concentration of binding (in such inter-molecular pathways) arises for a different physical reason\cite{thonhauser} 
than in chemical bonds. Chemical binding can exist in the 
$\lambda\to 0$ limit (in a Hartree-Fock description) by orbital hybridization, 
but that cannot happen for the nonlocal-correlation (or vdW) binding.  While the inclusion of the $E_{\rm c}^{\rm nl}$ energy term in DFT calculations 
causes density changes, and therefore an electrostatic  signature,\cite{thonhauser} 
the total nonlocal-correlation binding reflects, instead, an energy gain\cite{thonhauser} produced by collective exitations, i.e., by plasmons 
described by the screening properties.\cite{jerry65,ma,anlalu96,rydberg03p126402,Dion,Langreth05p599,hybesc14}

\begin{table}
\caption{Bond lengths $d$ and binding energies of noble-gas dimers and trimers, as computed in vdW-DF-cx for fully relaxed structures. Experimental reference values, Ref.~\onlinecite{Ogilvie92}, are listed in parenthesis when available. The middle column shows our results for the total binding energy, 
$\Delta E$, of the noble-gas complexes. The last last column shows the nonlocal correlation contribution to the binding, 
$\Delta E_{\rm c}^{\rm nl}$.
\label{tab:nobgas}
}
\begin{tabular}{lccccc}
\hline \hline
        & \multicolumn{2}{c}{$d$ (\rm{\AA})}  & \multicolumn{2}{c}{$\Delta E $ (meV)}  &  $\Delta E_{\rm c}^{\rm nl}$ (meV)  \\
\hline
        Ne dimer   &  3.09 & (3.09)  &   10.2  & (3.64)   &   14.0   \\
        Ne trimer  &  3.09 &         &   30.2  &          &   42.8   \\
        Ar dimer   &  3.99 & (3.76)  &   18.9  & (12.3)   &   25.7   \\
        Ar trimer  &  4.02 &         &   55.1  &          &   75.4   \\
        Kr dimer   &  4.33 & (4.01)  &   22.1  & (17.3)   &   30.7   \\
        Kr trimer  &  4.35 &         &   64.6  &          &   90.8   \\ 
\hline
\end{tabular}
\end{table}

\begin{figure}
\includegraphics[width=0.48\textwidth]{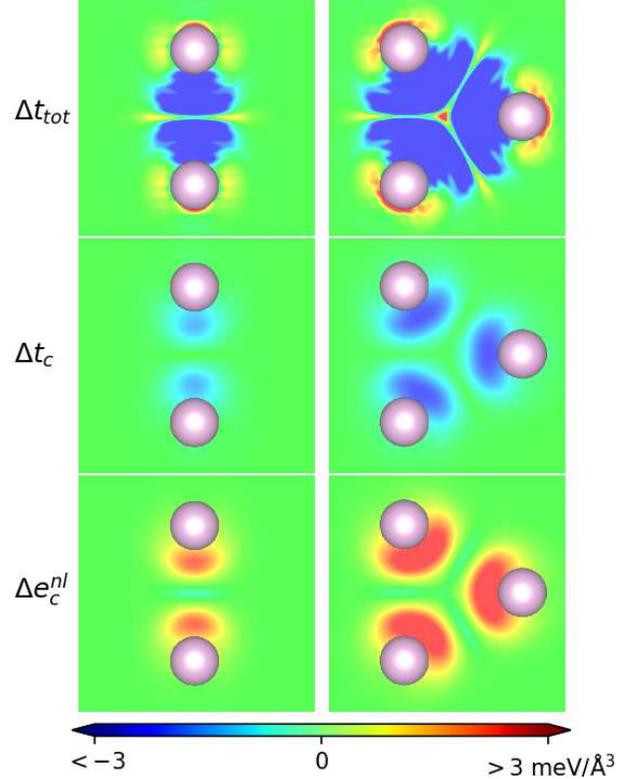}  
\caption{\label{fig:KRcomplex} Spatial variation in contributions to the
dispersion binding of the Kr dimer (left column) and of the Kr trimer (right column). 
The mapping is provided at the vdW-DF-cx results for the optimal structure, 
Table \ref{tab:nobgas}. The top, middle, and bottom pair of panels contrast the spatial
variation in the total kinetic energy $\Delta \langle \hat{T} \rangle$, 
the kinetic-correlation energy $\Delta T_{\rm c}$, and the nonlocal-correlation 
energy $\Delta E_{\rm c}^{\rm nl}$, respectively. The nonlocal part of the
kinetic-correlation energy, $\Delta t_{\rm c}^{\rm nl}(\vec{r})\approx 
\Delta t_{\rm c}(\vec{r})$, mirrors the variation in 
$\Delta e_{\rm c}^{\rm nl}(\vec{r})$.}
\end{figure}

\begin{figure*}
\centering
\includegraphics[width=0.8\textwidth]{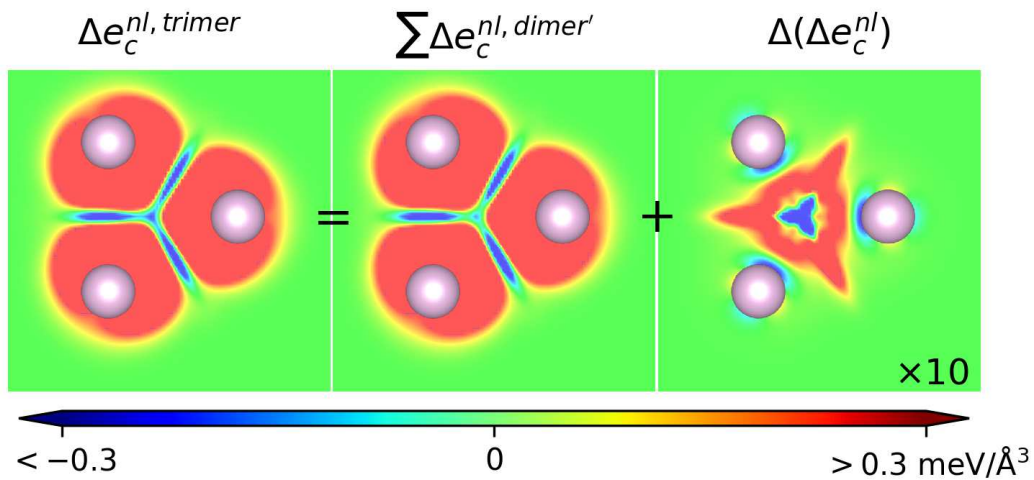}
\caption{Nonadditivity of the nonlocal correlation energy binding contributions in the Kr trimer. The left panel shows the nonlocal correlation contribution to the binding energy in the trimer, directly. The middle panel shows the results of simply making a superposition of the three dimer nonlocal correlation energy binding contributions (evaluated
with fixed bond lengths as set by vdW-DF-cx optimization of the Kr trimer). The right panel shows the spatial variation difference between the actual
trimer description and that of the dimer-based superposition, that is, 
$\Delta(\Delta e_{\rm c}^{\rm nl}) = \Delta e_{\rm c}^{\rm nl,trimer} - 
\sum \Delta e_{\rm c}^{\rm nl,dimer'}$.
\label{fig:nonadditive}
}
\end{figure*}

\subsection{Noble-gas complexes}

Figure \ref{fig:KRcomplex} contrasts contributions to the dispersion binding
in the Kr dimer (left column of panels) and the Kr trimer (right column of panels).
The results are presented for the optimal structure computed in vdW-DF-cx, 
Table \ref{tab:nobgas}, and for the vdW-DF-cx solution density. 

The top and middle rows of Fig.\ \ref{fig:KRcomplex} compare the 
spatial variations in the total kinetic energy and in the kinetic-correlation
energy. The KS kinetic energy effect makes up the larger part in the 
$\Delta t_{\rm tot}(\vec{r})$ variation because the dispersion interaction 
leads to density changes and thus to orbitals shifts. However, we find 
that there are also significant $\Delta t_{\rm c}(\vec{r})$ 
contributions arising between the noble-gas atoms. 

The variation in the $\Delta e_{\rm c}^{\rm nl}(\vec{r})$ binding contributions, 
bottom set of panel in Fig.\ \ref{fig:KRcomplex}, can again be used for a
qualitative discussion of the nature of binding. We find (as also documented for
the dispersion-bond cases investigated above) that the nonlocal part of the 
kinetic-correlation energy, $\Delta t_{\rm c}^{\rm nl}(\vec{r})$ (not shown), 
reflects the  $\Delta t_{\rm c}(\vec{r})$ variation (shown in the 
middle row of panels) and that $-\Delta t_{\rm c}^{\rm nl}(\vec{r})$ 
thus mirrors the variation $\Delta e_{\rm c}^{\rm nl}(\vec{r})$.

We make two observations about the nature of dispersion binding in
such noble gas complexes, based on Fig.\ \ref{fig:KRcomplex}. First, the
vdW-DF method describes the binding as arising in the region between 
(and not on) the noble-gas atoms. This observation is consistent with a previous 
vdW-DF characterization of the weak Ar$_2$ charge relocations that arise 
with the inclusion of the nonlocal correlation term $E_{\rm c}^{\rm nl}$, 
Ref.\ \onlinecite{thonhauser}. However, the vdW-DF-cx picture of dispersion 
binding is different from the London picture that suggests an atom-centred 
description.\cite{lo30,lo37}

Second, the vdW-DF-cx description of binding in the Kr trimer is not 
additive, i.e., the vdW-DF-cx account of the Kr trimer is not merely
a sum of dimer contributions. This is evident in Table \ref{tab:nobgas}
and it also holds when freezing the dimers at the slightly longer 
binding separations that characterize the Kr trimer.

Our discussion of nonadditivity is based directly on the
vdW-DF-cx account of dispersion interactions at binding separation,
Fig.\ \ref{fig:KRcomplex}. Nonadditivity of vdW forces\cite{barash,barash88,Sernelius98,Dobson06,kleis08p205422,Lebegue10,rev8,ts2,ts-mbd,DobsonGould12,TaoPerdew14,DobsonMB14,Dobson14,hybesc14,NanovdWScale} is often discussed in the context of a 
vdW description that is based on adding the series of 
asymptotic C$_6$, C$_8$, $\ldots$ coefficients.\cite{rev1,rev2,becke05p154101,becke07p154108,ts09,grimme3,Perdew12,PerdewTao12,Ruzsinszky12,Tao15,TaoRappe16}
However, we choose to instead utilize the fact that
our Kr dimer analysis provides us with a spatial
mapping of the vdW bond in a Kr dimer. This bond is 
located symmetrically around the axis between the two Kr atoms.
Comparing then such descriptions to that for the Kr trimer, we 
note a shift of the vdW binding towards the center region. 

Figure \ref{fig:nonadditive} provides details of this numerical exploration
of the vdW-DF-cx nonadditivity. The figure identifies where the 
trimer $\Delta e_{\rm c}^{\rm nl}$ variation differs 
from the variation in a sum-of-dimer-$\Delta e_{\rm c}^{\rm nl}$ 
description. Interestingly, the trimer binding changes 
(relative to  a sum of dimer contributions), arise in spatially 
confined pockets in the low-density, small density gradient regions.
As such, it further signals the importance of the inter-molecular
region (of sparse electron distribution) in the description of molecular 
and other sparse-matter binding.

\section{Summary and discussion}

We have provided formal analysis and calculations aiming to deepen the 
discussion of the nature of vdW interaction as described in the vdW-DF method.

A simple many-body physics effect underpins the vdW interaction, namely the 
mutual electrodynamical coupling of collective excitations (plamsons).\cite{jerry65,ma,ra,Dion,hybesc14}
This is an effect that exists in the fully interacting (physical) many-body  
system, described by coupling-constant value $\lambda=1$. The many-body physics 
effects manifest themselves both in the expectation value of 
the kinetic energy operator $\hat{T}$ and in the expectation value of the 
electron-electron interaction operator $\hat{V}$. However, in the standard -- and 
formally exact -- KS scheme for DFT calculations, we work with KS kinetic energy 
while incorporating the remainder, the kinetic-correlation energy
$T_{\rm c}$, within the formulation of an explicit XC functional $E_{\rm xc}$ 
(like PBE or vdW-DF-cx). 

We observe that a full characterization of the many-body physics effects behind the vdW interactions 
requires us to identify the XC contributions at $\lambda = 1$ and that such information
is available for the vdW-DF by a coupling-constant scaling analysis.\cite{Levy85,Levy91,Gorling93,Levy95chapter,Levy96,Perdew96,Burke97,Ernzerhof97}
The $\lambda = 1$ system is nominally given by an electron-gas response behavior 
that corresponds, instead, to the XC functional $E_{\rm xc}-T_{\rm c}$. This is one of
many consequences of the analysis presented in Refs.\ \onlinecite{Levy85,Gorling93,Burke97,Ernzerhof97}.

We present a code, called \textsc{ppACF}, so that we can extract spatially resolved 
binding contributions $\Delta e_{{\rm c},\lambda=1}^{\rm nl}(\vec{r})= \Delta e_{\rm c}^{\rm nl}(\vec{r}) - 
\Delta t_{\rm c}^{\rm nl}(\vec{r})$ for this description. We also provide this full-interaction
characterization of the nature of the vdW interaction mechanism\cite{jerry65,ra,ma,hybesc14} 
for intra-molecular binding, for typical inter-molecular binding 
cases of the S22 benchmark set,\cite{S22} and for a Kr cluster. 

Overall, our results for weakly bonded systems confirm 
that it is the sparse density gradient region between 
molecules which dominates the contributions to the vdW interactions.\cite{rydberg03p126402,kleis08p205422,berland10p134705,berland11p1800,rationalevdwdfBlugel12,behy13,behy14,hybesc14}  The vdW interaction is often perceived and handled as an atom-centered effect, 
i.e., consistent with the original London picture of vdW forces.\cite{lo30,lo37} 
However, as we have also illustrated here, the vdW-DF-cx calculations reveal a different picture. 

The vdW-DF ability to handle binding arising in electron 
tails\cite{ma,anlalu96,rydberg03p126402,hybesc14} is important, for it 
naturally leads to an enhancement of the interaction at binding distances. 
This is true even if the asymptotic dispersion forces may be  weak.\cite{Berland_2015:van_waals} 
In the case of extended systems this enhancement effect can be interpreted 
as image-plane effects at external or internal surfaces.\cite{zarembakohn1976,zarembakohn1977,harrisnordlander1984,rydberg03p126402,kleis08p205422,berland09p155431,berlandthesis,lee11p193408,lee12p424213,rationalevdwdfBlugel12,berlandthesis,behy13} 
For molecules, it is more natural to discuss the binding enhancement through the 
observation that it is much easier to polarize the electron distribution in the 
tails than in high-density regions near the atom nuclei.\cite{ra,ma,anlalu96,becke05p154101,becke07p154108,Dion,kleis08p205422,berland10p134705,berland11p1800,hybesc14} 

We also highlight that the nonlocal-correlation binding among molecules has 
signatures, channeled into pockets, i.e., concentrated in regions that resemble 
an orbital structure. The binding structure, as revealed in $\Delta e_{\rm c}^{\rm nl}(\vec{r})$  or in $\Delta e_{{\rm c},\lambda=1}^{\rm nl}(\vec{r})$, shares characteristics that resemble 
(but is much weaker than) those found in the KS kinetic-energy account of traditional chemical binding. 
This observation is useful for developing 
the qualitative discussions of vdW forces. For example, we can use such bond
signatures to document that the vdW-DF-cx account of the nonlocal-correlation 
binding in the Kr trimer is not additive -- the binding is not merely a sum 
nonlocal-correlation binding contributions in Kr dimers.

Finally, we note that our coupling constant scaling results allow us to 
generalize the construction of strictly parameter-free ACF-based 
hybrids\cite{Levy91,Gorling93,Perdew96,Burke97,Ernzerhof97} to a foundation in the vdW-DF method. Such constructions and the implications for the
use of the vdW-DF-cx0 hybrid\cite{DFcx02017} is presented
in a forthcoming paper.

\section*{\label{sec:ack} Acknowledgement}

We thank Kristian Berland and Jeffrey B.\ Neaton for useful discussions. Work supported by the Swedish Research Council (VR) through grants No. 2014-4310 and 2014-5289 and the Chalmers Area-of-Advance-Materials theory activity. The authors also acknowledge computer allocations from the Swedish National Infrastructure for Computing (SNIC), under contract 
SNIC2016-10-12 and SNIC 2017/1-174, and from the Chalmers Centre for Computing, Science and Engineering (C3SE) under 
contract C3SE2017-1-3.

\appendix

\section{Coupling constant scaling}

The coupling-constant result Eq.\ (\ref{eq:scaleExc}) can be obtained
from a renormalization-group perspective on the ACF. 

In the coupling-constant analysis of XC functionals\cite{Levy85,LYP85,Levy91,Gorling93,Levy96,Perdew96,Burke97,Ernzerhof97}
we consider a would-be many-body physics problem specified by the Hamiltonian
\begin{equation}
    \hat{H}_\alpha = \hat{T} + \alpha \hat{V} 
    + V_{\rm ext} \, ,
\end{equation}
where $\hat{V}$, again, denotes the operator for full (or actual)
electron-electron interaction, and where $V_{\rm ext}$ is the 
external potential, in part set by the nuclei. The formal machinery 
of DFT \cite{hoko64,kosh65} works for $\hat{H}_{\alpha}$ as well as
it does for the actual, physical problem defined by $\hat{H}=\hat{H}_{\alpha=1}$. In such a generalized-DFT
framework, we use $\Psi_n^{{\rm min},\alpha}$ to denote the 
ground-state many-body wavefunction solution which at any 
given $\alpha$ will be a unique functional of the density
variation $n(\vec{r})$, as indicated. Also, 
$\Psi_n^{{\rm min}}$ denotes the ground-state solution 
for the physical problem, at $\alpha=1$, i.e., the problem
for which we normally employ the DFT construction using a KS
calculational scheme.\cite{kosh65}

For any given $\gamma$ we can consider the density scaling
$n(\vec{r}) \to \gamma^3 \, n(\gamma \vec{r})$, where the choice
$\gamma=1/\lambda$ corresponds to the scaling that was discussed 
in the main text. The key observation\cite{Levy85} is that this scaling 
permits us to formally construct the coupling-constant scaling in the  ground-state many-body wavefunction solutions to $\hat{H}_{\alpha}$, 
\begin{equation}
    \Psi_{n}^{{\rm min},\alpha}(\vec{r}_1,\ldots, \vec{r}_N)
    = \alpha^{3N/2} \Psi_{n_\gamma}^{\rm min}(\alpha \vec{r}_{1}, \ldots, \alpha \vec{r}_{N}) \, ,
\end{equation}
for $\gamma=\alpha^{-1}$, Ref.\ \onlinecite{Levy91}. We note that $\Psi_{n_\gamma}^{\rm min}$ is a ground state solution for the 
density $n_{\gamma=1/\alpha}$. The KS kinetic-energy 
functional is defined\cite{Levy91}
\begin{equation}
T_{\rm KS} [n] = \langle 
\Phi_{n}^{{\rm min}} 
| \hat{T} |
\Phi_{n}^{{\rm min}} 
\rangle \, ,
\label{eq:KSFuncDef}
\end{equation}
where $\Phi_n^{\rm min}$ is a Slater-determinant wavefunction
that corresponds to the density variation $n(\vec{r})$ and 
minimizes the expectation value in Eq.\ (\ref{eq:KSFuncDef}).

For any assumed value of $\alpha$, a generalization of the XC standard energy functional,  
\begin{equation}
E_{\rm xc}^{\alpha}[n]\equiv \langle \Psi_n^{{\rm min},\alpha}
|(\hat{T}+\alpha \hat{V}) | \Psi_n^{{\rm min},\alpha} \rangle - T_{\rm KS}[n] - \alpha
U[n] \, ,
\label{eq:GenExcForm}
\end{equation} 
will permit us to pursue DFT calculations in a generalized KS scheme,
solving the $\hat{H}_{\alpha}$ problem.\cite{kosh65,lape77,LYP85,Burke97}
In Eq.\ (\ref{eq:GenExcForm}), $U[n]$ denotes the mean-field Coulomb interaction 
among electrons, Eq.\ (\ref{eq:Udef}). 
Similar as for the wavefunctions, we reserve the subscript-free
version $E_{\rm xc}[n]$ to denote the standard XC energy functional, 
i.e., relevant for the standard KS scheme in DFT.\cite{kosh65}

Following Refs.\ \onlinecite{Levy85,LYP85,Burke97}, we further define
$\alpha$-specific density functionals for the expectation values of the electron-electron interaction
\begin{equation}
V_{e-e}^{\alpha}[n] \equiv \langle 
\Psi_n^{{\rm min},\alpha} 
| \hat{V} |
\Psi_n^{{\rm min},\alpha} 
\rangle \, ,
\end{equation}
for expectation value of the kinetic-energy operator,
\begin{equation}
T^{\alpha}[n] \equiv \langle 
\Psi_n^{{\rm min},\alpha} 
| \hat{T} |
\Psi_n^{{\rm min},\alpha} 
\rangle \, .
\end{equation}
At any given density $n$, it follows that $\alpha U[n]
=\alpha^2 U[n_{1/\alpha}]$, $T_{\rm KS}[n]
=\alpha^2 T_{\rm KS}[n_{1/\alpha}]$ and that 
\begin{equation}
    T^\alpha[n] + \alpha V_{e-e}^{\alpha}[n] = \alpha^2
    \langle 
\Psi_{n_\gamma}^{{\rm min}} 
| (\hat{T} + \hat{V}) |
\Psi_{n_\gamma}^{{\rm min}} 
\rangle \, ,
\label{eq:scaleFunctalpha}
\end{equation}
for $\gamma=1/\alpha$. 

Next, we revisit the ACF,\cite{lape75,gulu76,lape77} Eq.\ (\ref{eq:ACF}), noting  that we might just as well use it to define and compute the 
$\alpha$-specific XC functional
\begin{equation}
E_{\rm xc}^{\alpha}[n]= \int_0^\alpha d\lambda \, E_{{\rm xc},\lambda}[n]
\, .
\label{eq:RGcondition}
\end{equation}
This XC functional can be used in a  
KS scheme for solving the $\hat{H}_\alpha$ ground-state problem.\cite{kosh65,lape77,Levy85,LYP85} The  density-scaling result 
Eq.\ (\ref{eq:scaleFunctalpha}) implies that such XC energy 
functionals adheres to a simple scaling result,\cite{Levy85,LYP85}
\begin{equation}
    E_{\rm xc}^{\alpha}[n] = \alpha^2 \, E_{\rm xc}[n_{1/\alpha}] \, .
\end{equation}

Finally, the main scaling result for the XC energy functional, Eq.\ (\ref{eq:scaleExc}), follows from the renormalization condition, Eq.\ (\ref{eq:RGcondition}), by simple derivation in the assumed value of
 the coupling constant in $\hat{H}_{\alpha}$.


%
\end{document}